\documentclass[nonblindrev]{informs4} 
\JOURNAL{ }
\DoubleSpacedXI 
\usepackage{natbib}
 \bibpunct[, ]{(}{)}{,}{a}{}{,}%
 \def\bibfont{\small}%
\usepackage{subcaption}
\usepackage{float}
\graphicspath{{picture/}}
\TheoremsNumberedThrough
\ECRepeatTheorems
\EquationsNumberedThrough
\MANUSCRIPTNO{-}
\usepackage{anyfontsize}
\begin{document}
\RUNAUTHOR{Zhu and Song}
\RUNTITLE{Information Disclosure in Auctions}

\TITLE{Design Information Disclosure under Bidder Heterogeneity in Online Advertising Auctions: Implications of Bid-Adherence Behavior}

\ARTICLEAUTHORS{%
 \AUTHOR{Mingxi Zhu}
 \AFF{Scheller College of Business, Georgia Institute of Technology, Atlanta, GA 30332, \EMAIL{mingxi.zhu@scheller.gatech.edu}}
 \AUTHOR{Michelle Song}
 \AFF{Google, 1600 Amphitheatre Parkway, Mountain View, California 94043, 
 \EMAIL{yingze@stanford.edu}}
 } 
\ABSTRACT{ Bidding is a fundamental aspect of search advertising today. It is less commonly recognized that bidders in online advertising differ both in their valuations and in their bidding strategies. Fortunately, disclosing bid information not only uncovers the heterogeneity in valuations and strategies among bidders, but also allows the platform to customize its information disclosure policies to achieve various objectives, such as enhancing consumer surplus, or increasing platform revenue. We analyzed data from a platform that implemented a system for bid recommendations based on disclosing historical bids. Our findings highlight the diversity in bidding strategies among advertisers: some directly adhere to the platform's disclosed historical bid information, while others construct their own unique bids that diverge from the platform's revealed data. This underscores the importance of designing information disclosure in online ad marketplaces in response to bidder diversity. We developed an equilibrium model for the Generalized Second Price (GSP) auction and demonstrated that adhering to any bid recommendations with a positive probability is theoretically suboptimal.  We then distinguished advertisers into categories of \textit{bid-adhering} and \textit{bid-constructing} bidders. We developed a novel structural model for self-bidding in GSP auctions to identify the private valuations of heterogeneous advertisers. This model facilitates a counterfactual analysis to assess the impact of varying levels of information disclosure on the market. Both theoretical and empirical evidence indicate that a moderate increase in information disclosure significantly enhances platform revenue and overall market efficiency. Identifying the diversity in bidder valuations and strategies is essential for online advertising platforms. By carefully designing information disclosure policies, platforms can not only identify heterogeneity among bidders to better address diverse bidder needs, but also efficiently achieve their managerial objectives through costless information disclosure.
}


\maketitle

 \section{Introduction}

Search advertising is a significant component of online advertising, with its revenue in the United States being projected to reach US \$ 132 billion dollars in 2024 (\cite{statista2024search}). Major platforms such as Google and Yahoo utilize the complicated Generalized Second Price (GSP) auction system. This system demands expertise in quality scores, bid distributions, and click-through rates at various positions -- knowledge often inaccessible to small business owners with limited resources. Such complexity not only highlights the challenges faced by advertisers, but also underscores the need for platforms to provide information and strategies that are more accessible to less experienced bidders.

To address the challenges faced by individual advertisers in the bidding process, advertising platforms have introduced several strategies to enhance understanding of GSP bidding system for bidders. One widely adopted strategy is auto-bidding. Auto-bidding allows advertisers to set their budgets, maximum bid limits, and targets for impressions, with the platform serving as a consumer agent. While this mechanism supports less experienced advertisers, as bidders do not have control over their own bids under auto-bidding system, it raises issues related to fairness and transparency (\cite{akbarpour2018credible,sej2023google}). This paper, on the other hand, shifts focus to another important aspect of digital advertising strategy: the role of information disclosure in self-bidding environments, a practice once adopted by Google and Yahoo, and currently in used by the platform we collaborated in this study. Self-bidding, empowered by information disclosure, allows advertisers to make informed decision and manually set their own bids, which enhances the overall transparency of system. Although analytical models suggest that platform can enhance its revenue through information disclosure (\cite{amaldoss2015keyword}), empirical evidence supporting the efficacy of these strategies is scant, primarily due to the lack of quasi-experimental data in self-bidding environments.

This paper examines how information disclosure affects advertiser behavior and market equilibrium. We address key questions: How does information release impact market outcomes? How do advertisers respond differently to disclosed information? How can disclosure strategies be improved on advertising platforms? To study these issues, we collaborate with a Yelp-like lifestyle service platform in China, which provides bid recommendations to wedding service advertisers in one city but not others. This quasi-experimental setup enables us to analyze the effects of information disclosure on advertising auctions, market outcomes, and advertiser behaviors.

In our paper, we identify a subset of bidders in GSP auctions who deviate from equilibrium bids by following the platform’s recommendations, which we show to be suboptimal. Detecting such behavior in real-world settings has been difficult due to the lack of quasi-experimental data. We classify these bidders as \textit{bid-adhering} and those who create their own bids as \textit{bid-constructing}. To analyze the impact on market welfare, we develop a novel structural model that identifies the private values of bid-constructing bidders and estimates value bounds for bid-adhering bidders. Finally, our counterfactual analysis shows that the platform can influence surplus outcomes by adjusting the level of information disclosure.

This article contributes to the literature on information disclosure by providing empirical, theoretical, methodological, and policy insights.

Empirically, we document varied advertiser responses to bid recommendations in the GSP auction, providing the first empirical evidence that some advertisers follow these recommendations even when it is suboptimal to do so. This underscores the critical impact of information disclosure policy on market dynamics through affecting heterogeneous groups of advertisers. 

Theoretically, we prove that bid-adhering bidders do not maximize their expected surplus by following bid recommendations. This suggests that the bid-adhering bidders are adopting different strategies from the bid-constructing ones, highlighting the  bidder diversity in bidding strategies under information disclosure. 

Methodologically, we develop a novel approach for valuation identification for GSP auctions, expanding on methods previously limited to first-price auctions (\cite{guerre2000optimal}). Our methodology differs from other studies by not relying heavily on assumptions about the specific bidding strategies of different bidder groups in equilibrium. This distinguishes it from approaches focused on valuation identification for asymmetric bidding strategies in auctions (e.g., \cite{krasnokutskaya2011identification}, \cite{hortaccsu2008understanding}, \cite{gayle2008numerical}). Such methodology could be readily adopted in other GSP auctions.  

On a policy level, this paper evaluates the impact of information disclosure on advertising platforms with diverse advertisers. The findings highlight the need to consider the varied responses of bidders when designing information-disclosure policies for online advertising. We demonstrate that through designing information disclosure, the platform can significantly enhance both ads revenue and market efficiency.

The paper is organized as follows: Section 2 reviews the related literature. Section 3 provides the empirical background and details the information disclosure event. Section 4 presents evidence of advertisers' varied responses to information disclosure and discusses the suboptimality of bid-adhering strategies. Section 5 introduces our empirical model for GSP auctions, including valuation identification and estimation results. Section 6 conducts a counterfactual analysis on how changes in information disclosure affect market outcomes. Section 7 concludes with a summary and suggestions for future research.

\section{Literature Review}

Our work relates to several streams of literature: the impact of information disclosure on market outcomes, GSP auctions for online advertising, empirical and theoretical studies of online advertising, the interaction between human behavior and platform policies, and methodologies for valuation estimation. We review each of these areas separately below.

\textbf{Impact of Information Disclosure on Market Outcomes}: This study contributes to the literature on the impact of information disclosure on market outcomes. In the supply chain context, the effects of information sharing vary: \cite{chu2017strategic} and \cite{cui2018sharing} find that, under certain conditions, carefully managed information sharing can benefit supply chain participants, while \cite{shamir2018perils} and \cite{lu2023forewarned} show that accurate information sharing is not always beneficial and may even harm participants. In crowdfunding platforms, \cite{chakraborty2021signaling} demonstrates through a game-theoretic model that well-designed information signals can enhance platform performance. In online matching platforms, \cite{chen2024signaling} finds that information signaling improves matchmaking by dynamically balancing supply and demand. In auctions, \cite{tadelis2015information} shows that in wholesale automobile auctions, providing quality information enhances the match between dealers and cars boosts auction revenue. In our study, we find that information disclosure enhances both platform revenue and overall surplus in online advertising markets by encouraging bid-adhering bidders to bid more effectively.

\textbf{GSP Auctions}: Since Google introduced the GSP auction in 2002, it has been widely studied in the context of search advertising. \cite{edelman2007internet} and \cite{varian2007position} develop the envy-free equilibrium model, which minimizes platform revenue but maximizes advertiser surplus under complete information. \cite{athey2010structural} further extends this model by incorporating dynamics more representative of real-world GSP auctions. While these studies assume rational, Bayesian utility-maximizing bidders, our work identifies bid-adhering behavior in a subgroup of bidders that cannot be explained under this assumption. This highlights the need for a new empirical model for valuation identification in online search advertising with heterogeneous bidders.

\textbf{Empirical and Theoretical Studies of Online Advertising}:, \cite{choi2020online} and \cite{agrawal2020optimization} provide comprehensive literature reviews and commentaries on online and digital advertising markets. Empirically, \cite{jerath2011position} shows that under a pay-per-click mechanism in GSP auctions, less reputable firms have strong incentives to bid aggressively. Similarly, \cite{narayanan2015position} finds that position effects are more pronounced for smaller advertisers and when consumers are unfamiliar with the advertiser’s keyword. Other studies, such as \cite{ghose2009empirical}, \cite{zhang2011cyclical}, \cite{yao2011dynamic}, and \cite{agarwal2011location}, focus on  consumer behavior and its impact on advertisers’ strategies. Theoretically, a stream of work focuses on providing theories for auto-bidding environment (\cite{balseiro2019learning}). In platform display decisions, \cite{najafi2014cost} and \cite{fridgeirsdottir2018cost} formulate the platform display problem in online advertising as a queueing system and design the optimal pricing schemes on both cost-per-click and cost-per-impression. Our work extends this stream of literature by incorporating information disclosure into the analysis, where we explore heterogeneous responses from bidders under such information disclosure.

\textbf{Interaction between Human Behavior and Platform Policies}: This paper contributes to the literature on the interaction between market participants' behaviors and platform mechanisms. In auction platforms, participant behaviors such as "sniping" in eBay auctions (\cite{ockenfels2006late, roth2002last, bajari2003winner, barbaro2004shilling}) and inefficient bidding in electricity auctions (\cite{hortaccsu2008understanding}) have been well-documented. \cite{hortaccsu2019does} studies how heterogeneity in firms sophistication level impact electricity market outcomes. \cite{caro2023believing} focuses on price-adhering behavior in decision support systems and identifies factors contributing to non-adherence. In the online real estate market, \cite{fu2022does} examines how information disclosure, such as platform home value estimates, influences human decisions and market outcomes. Our study identifies a new non-strategic behavior—adherence to bid recommendations—based on empirical evidence from real-world settings. We analyze the impact of this behavior on market outcomes and propose policy designs to improve revenue and social surplus.

\textbf{Methodology for Valuation Estimation}: This paper also contributes to the methodological literature on valuation estimation in online platforms. Early works on online auctions, such as \cite{donald1993piecewise}, \cite{donald1996identification}, and \cite{laffont1995econometrics}, focus on parametric identification of bidders' private values in first-price auctions with symmetric bidders. \cite{guerre2000optimal} (GPV hereafter) introduces a computationally feasible method for first-price auctions under the symmetric independent private value paradigm. \cite{campo2003asymmetry} extends this to asymmetric bidders with affiliated private values. In procurement auctions, \cite{kim2014measuring} develops a structural estimation approach to estimate bidders' private costs using bidding data from combinatorial auctions, building on the methods in \cite{cantillon2006auctioning} based on GPV. Our paper extends GPV methods for online Generalized Second Price (GSP) auctions, assuming a subset of bidders follows equilibrium strategies. Similar to GPV, our model infers bidder values from empirical bid distributions by calculating a positive markup term in GSP auctions. Additionally, many empirical studies test the Linkage Principle (\cite{milgrom1982theory}) in auctions, such as \cite{de2008impact}, \cite{cho2014linkage}, and \cite{gupta2019local}. However, in our case, there is no common value component in the information structure. In online retailing, \cite{jiang2024intertemporal} provides a nonparametric estimation method to detect consumer heterogeneity through transaction data. In our work, we use bid data to confirm heterogeneity in both bidder valuations and bidding strategies.
\section{Data and Background}
 
\subsection{The Ad Platform and the Generalized Second Price Auction}

\begin{figure}[htb!] 
    \centering
\includegraphics[height=.45\textwidth]{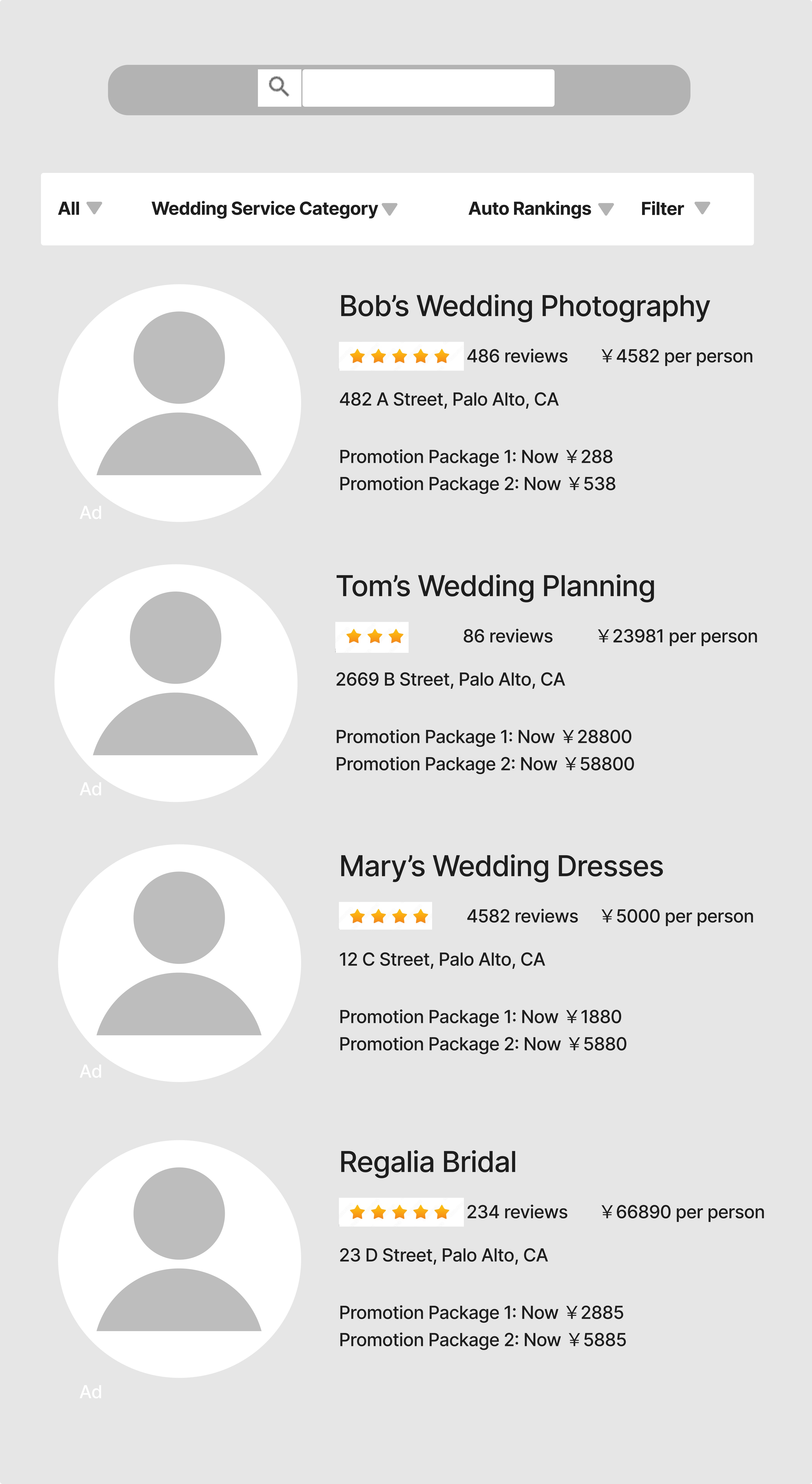}
\caption{Wedding Service Category Search Result App Layout}
	\label{fig:searchpage}
\end{figure}

We partnered with a leading online platform for discovering lifestyle services in China. The platform is similar to Yelp and Groupon in North America. As of 2023, it has 677 million annual transacting users and 9.3 million annual active merchants. The advertising platform manages sponsored search ads for various lifestyle services such as restaurants, takeouts, travel, and wedding services. Our research focuses on the wedding service category, where the platform has introduced bid recommendations as a means of disclosing information to advertisers. Figure \ref{fig:searchpage} displays the search result for wedding service providers on its mobile App. The top five slots on the search results page display advertisements marked with an ``ad" logo on the merchant's picture, which may be easily overlooked unless consumers closely inspect the page. Alongside these ads, the search results page also provides additional information about each merchant, including their title, review stars, number of reviews, average price per person, and location.

\vspace{-1em}
On this platform, the first five positions are reserved for search advertisements. Subsequent advertising results are mixed with organic search results, appearing every four spots starting from the ninth position (e.g., the sixth ad appears in the ninth slot). Advertisers bid for desired time slots to display their ads at the top of the page or interspersed with organic search results.  Advertisers bid on their targeted time periods to display their ads for sponsored search. They submit a single bid that applies to all positions, and the ad platform then ranks them for each search query based on a combination of their bids and quality scores. Specifically, when a consumer searches for a relevant keyword, the platform calculates the advertiser’s quality score for that specific consumer query, and ranks the advertisement based on a rank score, which is the product of the quality score and the submitted bid.

Like other sponsored search platforms, this ad platform operates using the GSP auction model (\cite{edelman2007internet}). Each consumer search initiates an auction, where the platform identifies matching advertisers, retrieves their current bids $b$, and assigns a quality score $q$ for that specific consumer search query. Advertisers are ranked by a rank score $r=q\times b$. If advertiser $i$ has the $k^{th}$ highest rank score, she occupies slot $k$. The pay-per-click price for ad slot $k$ to advertiser $i$, $p_{i,k}$ is determined by dividing the $k+1^{th}$ highest rank score $r_{k+1}$ by the advertiser $i$ 's quality score $q_{i}$, $p_{i,k} = \frac{r_{k+1}}{q_{i}}$. If the number of clicks (or in other words, the Click Through Rate, CTR) for position $k$ is $\alpha_k$, the advertiser $i$'s total payment for ad slot $k$ is $\alpha_k\times p_{i,k}  = \frac{\alpha_k r_{k+1}}{q_{i}}$. 

\subsection{Information Disclosure in Online Advertising Auctions}
\label{subsec:info_disclosure_mechanism}
The ad platform initially launched bid recommendations as a means of information disclosure for wedding service advertisers in Xi'an, China, on August $13^{th}$, $2019$, without extending this to other cities. Subsequently, on October $10^{th}$, $2019$, it expanded these recommendations nationwide. For our analysis, we selected Chengdu as a control city to assess the impact of these information disclosures in Xi'an using the difference-in-differences method. Chengdu was selected based on a weighted comparison of average bid, cost-per-click (CPC), number of ads, and advertising revenue during July 2019, a month before the policy change. Among 145 cities, Chengdu was the most similar city to Xi'an in these metrics. A robustness check with Nanjing, the second-best match, confirms that our findings are consistently valid.

The advertising platform updates bid recommendations daily, based on yesterday's second to fourth highest rank scores adjusted by each advertiser's simulated quality score, following a methodology similar to Google's first page bid estimates strategy. For instance, the bid recommendation for the first slot for advertiser \( i \), denoted as \( s_{i,1} \), is calculated by dividing yesterday's second-highest rank score \( r_2 \) by the quality score \( q_i \) of advertiser \( i \) (\( s_{i,1} = \frac{r_2}{q_i} \)). Intuitively, \( s_{i,1} \) represents the minimum bid required for advertiser \( i \) to secure slot 1, assuming all other bids remain as yesterday, and quality scores remain as simulated.

Regarding the quality score, since the actual score is computed for each consumer query, the platform needs to simulate the quality score $q_i$ for each advertiser $i$ when constructing bid recommendations. Specifically, the platform simulates a few consumers from various locations in the city, predicts the quality scores $q_i$ using a machine learning algorithm, and averages all the predicted quality scores to obtain an estimate of advertiser $j$'s quality score used in the bid recommendation. Advertisers receive only the bid recommendations, not the detailed historical bid or quality score data.

Considering that bid recommendations utilize all \textit{ex-ante} information, they cannot guarantee $100\%$ success in securing the top three positions when the advertiser chooses to adhere to the recommendations. This is because the bidding environment, including bids from other advertisers, consumer search queries, and quality scores, is constantly changing. It is important to note that the platform publicly discloses this mechanism, and all advertisers are fully aware of how the bid recommendations are constructed. This transparency makes it \textit{common knowledge} that following the bid recommendations for a specific position does not guarantee securing that position in the actual auction.

\begin{figure}[htb!] 
    \centering
	\includegraphics[width=.5\textwidth]{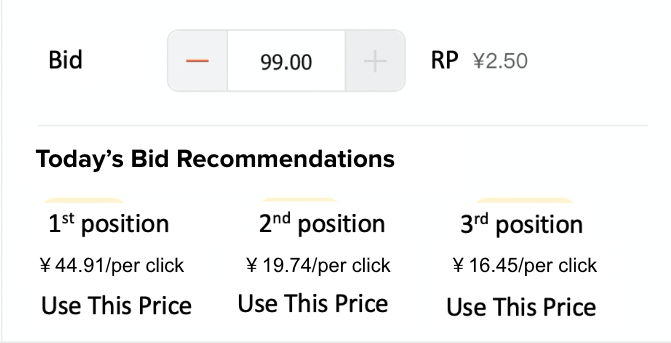}
	\caption{Advertiser's Bidding Interface}
	\label{fig:bidpage}
 \vspace{-2em}
\end{figure}

Figure \ref{fig:bidpage} displays the advertiser's bidding interface. ``RP'' stands for the reserved price. Below the bid submission area, advertisers can view the current day's bid recommendations for the first, second, and third positions. Beneath these recommendations, there is a button labeled ``Use This Price". When clicked, this button automatically sets the advertiser’s bid to the selected recommendation.

\subsection{Data}
The dataset spans from July $13^{th}$ to October $10^{th}$, $2019$. We analyze data from $30$ days prior to the introduction of bid recommendations on August $13^{th}$, $2019$, and $59$ days following the introduction. There are two primary datasets. The first dataset includes advertisers' performance level data, such as average daily Cost Per Click (CPC, in RMB), total daily number of clicks or Click Through Rate (CTR), the average quality score ({Avg\_qs}), and the average displayed position ({Avg\_pos}, indexed from $1$). The second dataset focuses on bid-change event, and contains fewer observations as it only includes the treated city post-intervention data. It records advertisers' activities when they log into their personal ad accounts, including any bid changes during the login, the original bid, the new bid, and the current displayed bid recommendations for the top three positions. Table \ref{sum_table_tex} provides summary statistics for these datasets, which will be used in our analysis. The CPC, bids and recommendations are in unit of RMB. Additionally, we have access to the advertiser review characteristics, including total number of reviews, and the good review ratio.

\begin{table}[htb!]
\begin{center}
{\small 
\def\sym#1{\ifmmode^{#1}\else\(^{#1}\)\fi}
\begin{tabular}{|llllll||llllll|}
\hline\hline
\multicolumn{6}{|l||}{\textit{Advertiser's Performance Level Data}} & \multicolumn{6}{l|}{\textit{Bid-change Event Level Data}} \\ \hline
             & \textbf{N}          & \textbf{Mean}    & \textbf{St. dev.}     & \textbf{Min.}      & \textbf{Max.}      &        & \textbf{N}       & \textbf{Mean}   & \textbf{St. dev.}   & \textbf{Min.}   & \textbf{Max.}   \\ \hline

CPC       & 73271      &  3.97     &  5.29  & 0.03  &  116.63  & Bids   & 3532   & 22.62 &21.16 &  1.00 &    99.00   \\
CTR          & 73271      &   75.88  & 10.33  &  54.17   &  102.69 & $s_1$      & 3532   &  54.62& 32.46&2.79& 99.00    \\
Avg\_qs       & 73271      &  0.12  & 0.05    &  0.02     &  0.47 & $s_2$      & 3532   & 41.87 & 26.21 & 2.11 & 99.00     \\
Avg\_pos      & 73271      &   35.67 & 28.29  &  1.04 &  455.48  & $s_3$    & 3532   &  33.50 & 22.15&1.00& 99.00\\
\hline\hline
\end{tabular}
}

\end{center}
\vspace{1em}
\caption {Summary Statistics Table}
\centering
\label{sum_table_tex}
\end{table}

 \section{Descriptive Evidence}\label{sec:descriptive}

\subsection{Documentation of Theoretically Sub-Optimal Bidding Pattern}
\begin{figure}
    \centering
    \includegraphics[width=0.6\textwidth]{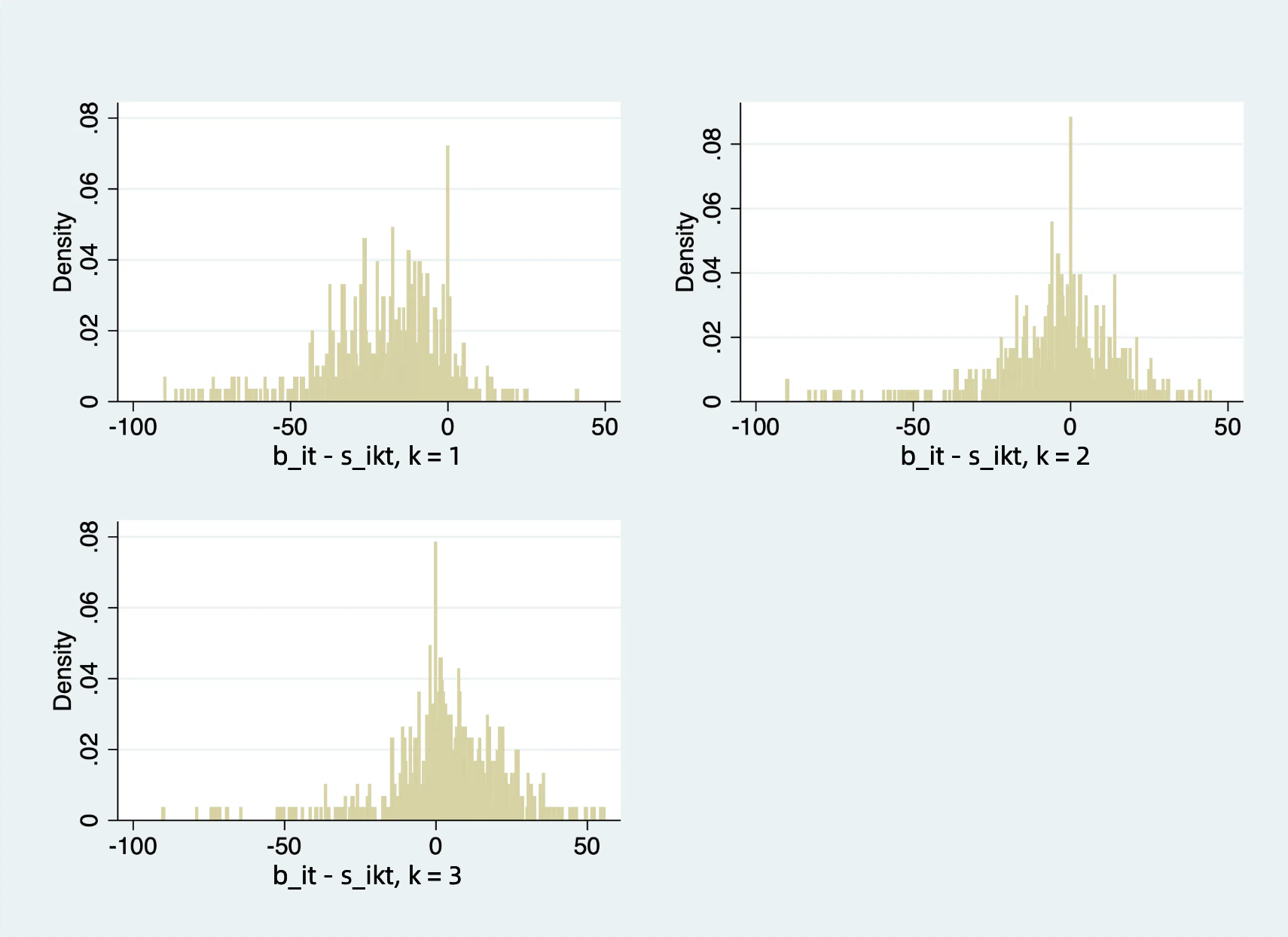}
\caption{\centering Bidding Pattern Under Information Disclosure}
\label{fig:naive}
\end{figure}

On August $13^{th}$, $2019$, the ad platform introduced bid recommendations for the top three ad positions to advertisers in the wedding service category in Xi'an. Figure \ref{fig:naive} illustrates the distribution of \(b_{i,t} - s_{i,k,t}\), where \(i\) indexes the advertiser, \(t\) indexes the bid change occasion, and \(k\) indexes ad positions. Here, \(b_{i,t}\) represents the bid submitted by advertiser \(i\) at time \(t\), and \(s_{i,k,t}\) is the current \(k^{th}\) position bid recommendation for advertiser \(i\) at time \(t\).  Figure \ref{fig:naive} illustrates the difference between each advertiser's submitted bid \(b_{i,t}\) and the three personalized bid recommendations \(s_{i,k,t}\) for the top $3$ positions, adjusted by quality scores, at each bid change occasion. The clustering around zero in Figure \ref{fig:naive} shows that advertisers frequently place bids exactly as recommended. In contrast, no similar pattern is observed in Chengdu, where we constructed pseudo-bid recommendations and compared them with advertisers' bids.

From Figure \ref{fig:naive}, we observe bid-adherence behavior under information disclosure, where some bidders directly follow the platform's recommended bids instead of constructing their own. The following theorem demonstrates that under mild assumptions, the strategy of following bid recommendation for any fixed position $k$ with strictly positive probability can not be rationalized. 

\begin{theorem}
\label{theorem:irrational}
Under Assumption \ref{assumption:irrational} (as detailed in Online Appendix \ref{irrational_proof}), if other bidders adhere to the bid recommendation for any fixed position \( k \), then any strategy that involves bidder \( i \) following the bid recommendation \( s_{i,k} \) for position \( k \) with strictly positive probability is not a best-response strategy.
\end{theorem}

The proof of Theorem \ref{theorem:irrational} 
is provided in Online Appendix \ref{irrational_proof}. And the intuition behind Theorem \ref{theorem:irrational} is that if the bidder is a Bayesian utility maximizer, she must submit a unique bid that optimizes her expected payoff, taking into account the bid distributions and quality score distributions of other bidders. Hence, the probability that this unique bid precisely matches the bid suggestion provided by the platform under the described mechanism is zero. Consequently, if a bidder adopts a strategy involving a positive probability of following the bid recommendation, such a strategy is not a best-response strategy. The bidder could benefit from reducing the probability of bid-adherence and submitting an optimally constructed bid instead. Hence, even a \textbf{\textit{one-time adherence}} to the bid recommendation for any position \(k\) cannot be rationalized. This allows us to further identify and distinguish bid-adhering bidders in our dataset as those who have placed bids matching the bid recommendations at least once.

Figure \ref{fig:naive} illustrates that bid-adhering behavior persists, even though it is theoretically sub-optimal. While some bidders follow bid recommendations with positive probability, others never bid at the cut-off and instead construct their own bids. In the next section, we categorize the bidders into two groups: \textit{bid-adhering} and \textit{bid-constructing} bidders, and delve into the descriptive patterns of bid-adhering bidders. 

\subsection{Descriptive Patterns and Explanations for Bid-adhering Bidders}
\label{subsec:descriptive_explain}
If adhering precisely to the bid recommendation is not theoretically optimal, why do advertisers still follow it? In this subsection, we provide descriptive patterns of bid-adhering bidders, and propose several mechanisms that could explain these theoretically sub-optimal bid-adhering behaviors.

Firstly, we define the elite
\textit{bid-adhering} bidders as those who have followed one of the three bid recommendations at least once. According to Theorem \ref{theorem:irrational}, adhering to the bid recommendation with any positive probability is not rational. The recommendations in our dataset are relatively high, indicating that these elite bidders have high valuations for online advertising. Although the group of potential bid-adhering bidders may be larger, we focus on elite bidders due to their direct observable behavior in our dataset and significant impact on platform revenue. In contrast, \textit{bid-constructing} bidders are those who have bid above the lowest recommendation at least once but have never exactly followed any of the recommendations. By requiring bid-constructing bidders to place bids above the lowest recommendation, we ensure that they have similar private valuations to the elite bid-adhering bidders.

In our data set, among the bid-adhering bidders, there are \(88\) individuals with a total of \(1494\) bid change event observations post-information release, where they bid at the cutoffs approximately \(13.5\%\) of the time. The group of bid-constructing bidders comprises \(99\) advertisers, with \(1101\) total bid change occasion observations post-intervention. The table below shows frequency statistics for bid-adhering bidders in online advertising, focusing on 18 active participants. Bidders with fewer than 10 bid changes are excluded to accurately represent frequent bidders.

\begin{table}[htb!]
\begin{center}
{\small 
\begin{tabular}{lccccc}
\hline
\textbf{Metric} & \textbf{N} &  \textbf{Mean } & \textbf{St. dev } & \textbf{Min} &  \textbf{Max } \\
\hline
Bid-Adherence Occurrences & 18 & 2.87 & 2.20 & 1 &9  \\
Total Bid Change Events & 18 & 28.80 &22.43 & 11 &103  \\
Bid-Adherence Rate $(\%)$ & 18 & $12.11\%$ & $10.02\%$ & $3.23\%$ &$41.67\%$  \\
\hline
\end{tabular}
}

\end{center}
\vspace{1em}
\caption {Bid-adhering Frequency Summary Statistics}
\label{table:bid_adhere_frequency}
\vspace{-1em}
\end{table}

Table \ref{table:bid_adhere_frequency} suggests that bid-adherence is a significant and substantive component of bidding behavior on the online-advertising platform, influencing not only the individual bidders outcomes but also the overall market equilibrium and efficiency. While the table underscores the prevalence of bid-adherence, it invites further exploration into the reasons why some bidders follow recommendations, whereas others, like bid-constructing bidders, adopt a different bidding strategy of crafting their own bids after information disclosure. Next, we analyze the differences between these two groups based on their bids, and propose several underlying mechanisms to explain why bid-adhering bidders follow the bid recommendations. 

\begin{figure}[htb!]
\centering
\begin{subfigure}{.5\textwidth}
  \centering
  \includegraphics[width=.75\linewidth]{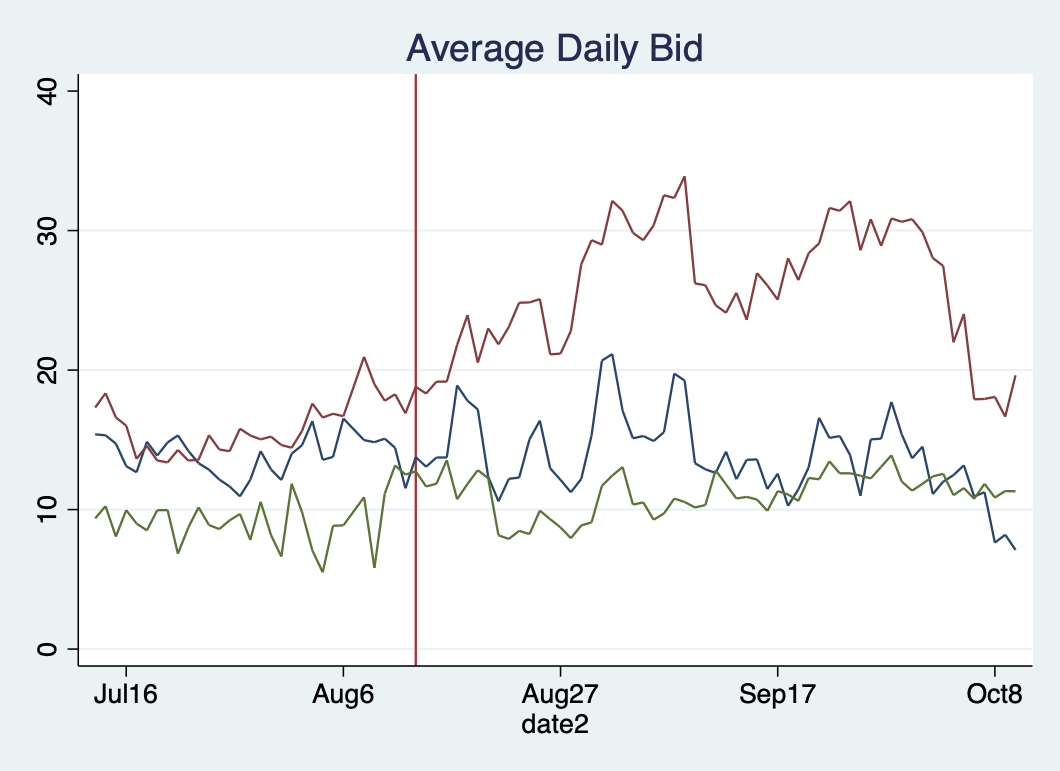}
  \label{fig:avg_bid}
\end{subfigure}
\caption{\centering Average Daily Bid across Bidders}

\footnotesize\textsuperscript{*} Figure \ref{fig:adinvestment} displays the average daily bid for bid-constructing bidders (shown in blue), bid-adhering bidders (shown in red), and bidders in the control city among top $25^{th}$-position advertisers (shown in green). The vertical line reflects the intervention date.
\label{fig:adinvestment}
\end{figure}

Figure \ref{fig:adinvestment} shows the average daily bid before and after the intervention. Initially, bid-adhering bidders submitted bids at levels similar to bid-constructing bidders. After the intervention, there was a significant increase in bids among the bid-adhering group, reflecting their adherence to higher recommendations for the top 3 slots. Conversely, bid levels among the bid-constructing group in the treated city and the top bidders in the control city remained constant. Figure \ref{fig:adinvestment} highlights the heterogeneity in bidding strategies between the groups after information disclosure intervention. Notably, there is a decrease in daily bids around early October. This is because early October is the peak season for wedding services in Xi'an, which necessitates advertising several days in advance, making late September a peak, and early October a trough for wedding service advertising with a significant drop on bids across groups. In Online Appendix \ref{app:review}, we also show that bid-adhering bidders differ from bid-constructing bidders in multiple characteristics according to our dataset, including their reviews and review stars.

Lastly, we present several potential mechanisms to explain bid-adherence and discuss the likelihood of these mechanisms. Due to the absence of laboratory data, we cannot precisely determine the exact reason for bid-adherence. However, from the support of descriptive patterns, we find that some explanatory mechanisms are more prevalent than others.

\textit{\textbf{Cognitive Ease}} : People often prefer to exert less cognitive effort, which makes them more likely to follow straightforward recommendations (\cite{sweller1988cognitive,jin2017different}). Our data indicates that bid-adhering bidders have a relatively high number of reviews but lower review stars (Online Appendix \ref{app:review}), suggesting they favor a straightforward volume-first strategy. This predisposition also makes them more likely to adopt straightforward recommendations for cognitive ease. Moreover, as the options of ``Use this Price" for bid recommendations are presented as default (Figure \ref{fig:bidpage}), they are more likely to be chosen as changing from a default option requires additional cognitive effort.

\textit{\textbf{Anchoring Effect}} : The anchoring effect describes how individuals may depend excessively on an initial piece of information (the ``anchor") for subsequent decision-making. Unlike the cognitive ease effect where bidders simply follow recommendations, the anchoring effect leads bidders to initially try to determine their optimal bid but eventually revert to recommended bids. Our descriptive evidences on reviews and review stars (provided in Online Appendix \ref{app:review}) suggest that bid-adhering bidders generally employ traditional marketing strategies aimed at increasing transaction volumes, reflecting a lower proficiency with advanced digital marketing and keyword search strategies. And those with less experience or knowledge in a domain are more susceptible to anchoring (\cite{wilson1996new}), offering a likely explanation for the observed bid-adherence.

\textit{\textbf{Learning Effect}} : Bidders might follow the bid recommendation to learn quality scores or top slot rank distributions, but we conjecture this effect to be limited since recommendations are observed prior to bidding. If the goal is to learn the bid or quality score distributions, advertisers can initially observe the recommendations and bid optimally without strictly adhering to these recommendations, then monitor outcomes to validate their beliefs, since the information disclosure occurs prior to their bidding. While learning effect is possible in auctions, such learning requires a stable environment and a relatively long-run play in a stable environment (\cite{doraszelski2018just}). As there is no clear and direct evidence of learning effects in our dataset, we decide to focus on short-term interactions under platform-driven changes in information disclosure policy, without considering learning dynamics.

\subsection{Impact of Information Disclosure On Market Outcomes}

\begin{figure}[htb!]
\centering
\begin{subfigure}{.5\textwidth}
  \centering
  \includegraphics[width=.75\linewidth]{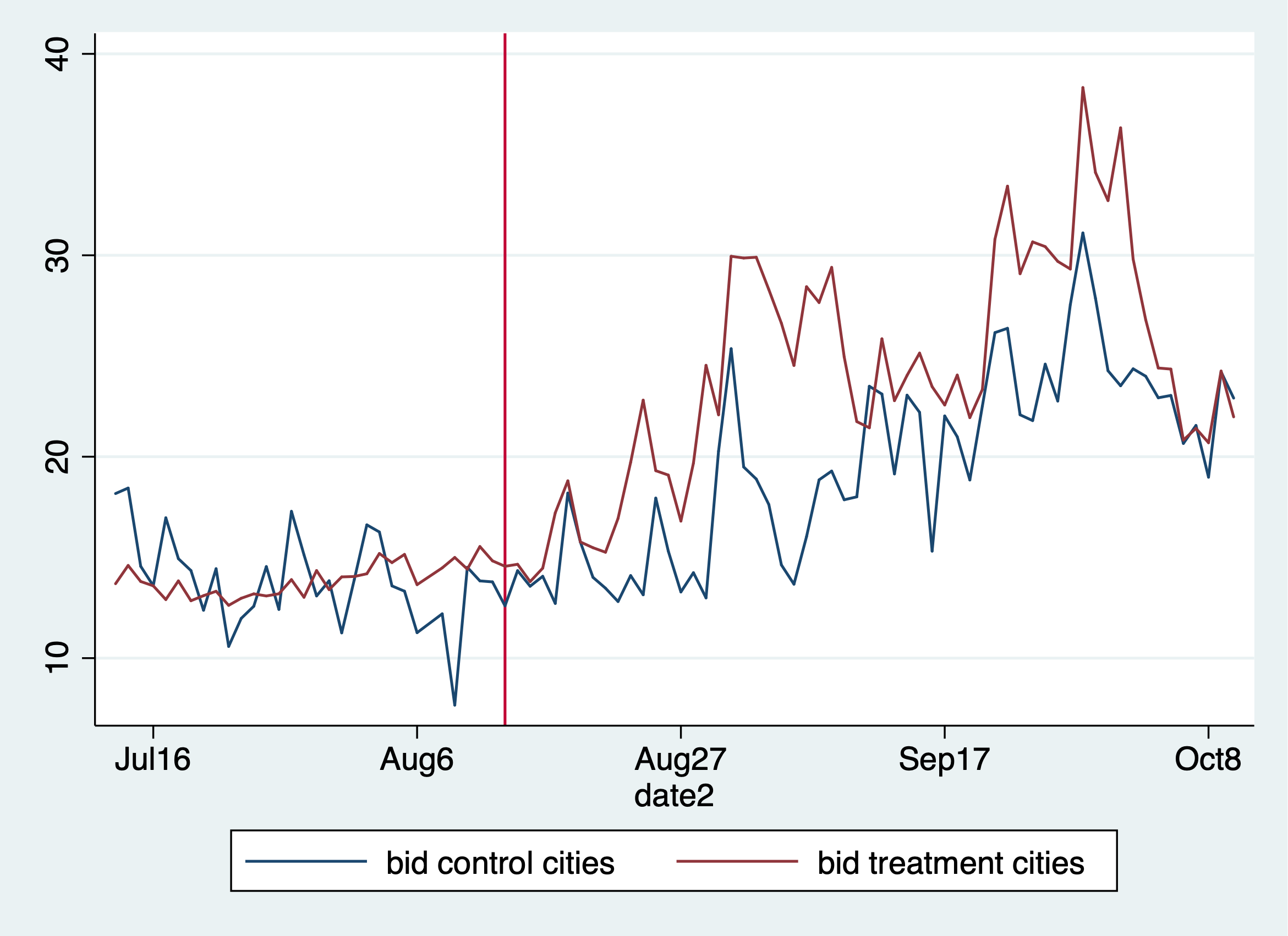}
  \captionsetup{font=footnotesize}
  \caption{Average Bid of Top 5 Ad positions}
  \footnotesize\textsuperscript{*} Figure \ref{fig:bid} displays the average daily bid level in Xi'an (red line) and Chengdu (blue line). 
  \label{fig:bid}
\end{subfigure}%
\begin{subfigure}{.5\textwidth}
  \centering
  \includegraphics[width=.75\linewidth]{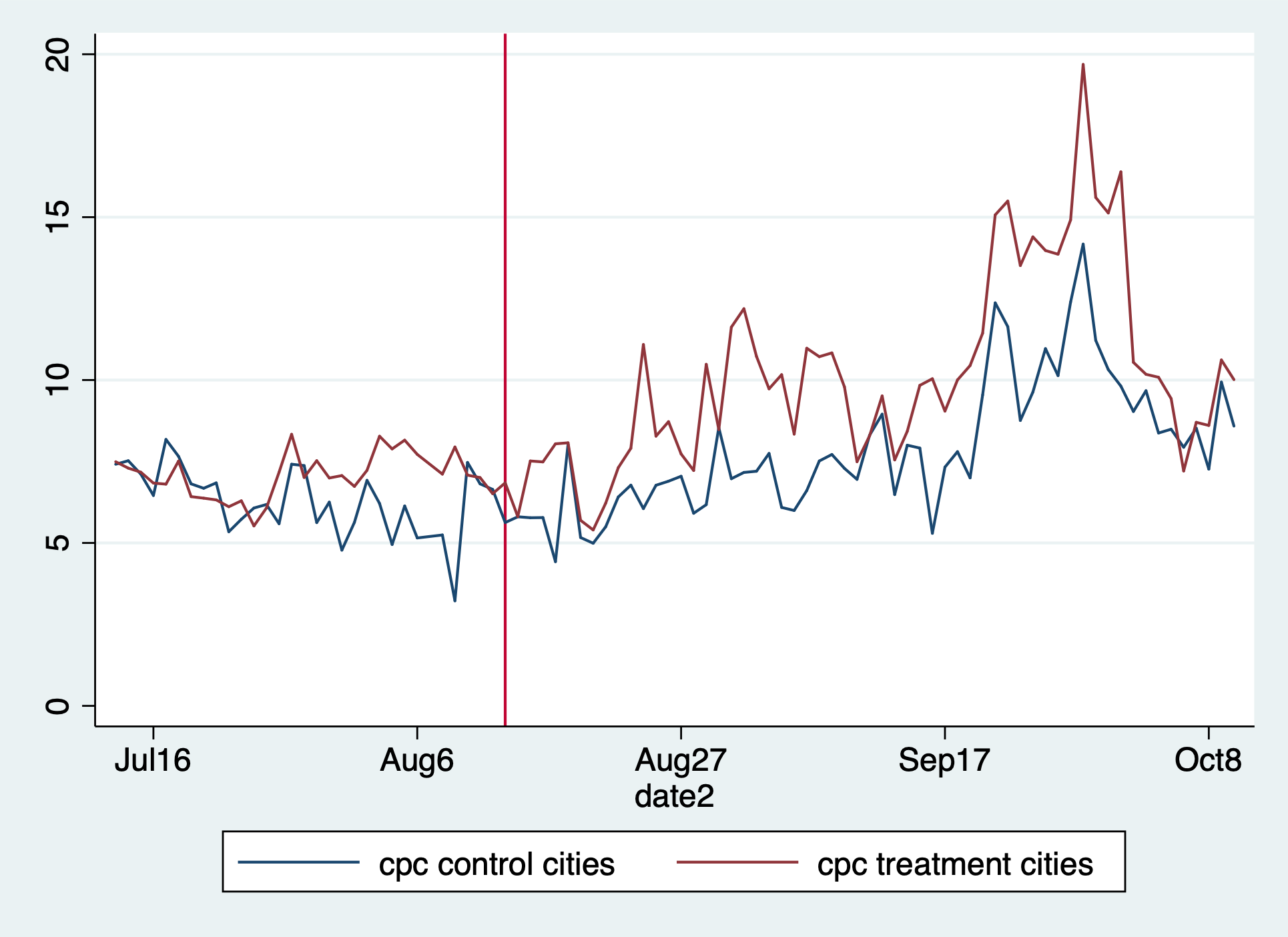}
\captionsetup{font=footnotesize}
  \caption{Average CPC of Top 5 Ad positions}
  \footnotesize\textsuperscript{*} Figure \ref{fig:cpc} displays the average daily CPC level in Xi'an (red line) and Chengdu (blue line). 
  \label{fig:cpc}
\end{subfigure}
\caption{\centering Pre-and-Post Changes in Bid and CPC of Top 5 Ad positions}
\label{fig:market outcome}
\vspace{-2em}
\end{figure}

In this section, we explore the impact of bid recommendation information disclosure on market outcomes. We first demonstrate the effects of bid recommendations on bids and cost per click (CPC) for the top five positions, as shown in Figure \ref{fig:bid} and \ref{fig:cpc}. The left graph presents the average daily bids for the top five ad positions in Xi'an (red line) and Chengdu (blue line) before and after the bid recommendation was introduced. The right graph displays the average CPC for the top five ad positions in both cities. Additional effects on lower positions are detailed in Online Appendix \ref{lowerplots}.

Before the policy change, average bids and CPC for top positions were similar in Xi'an and Chengdu. Following the implementation of bid recommendations on August 13, 2019, there was no immediate increase in the average bids and CPC in Xi'an compared to Chengdu. The divergence in bid levels became noticeable three days later, likely because advertisers may require time to adapt to this new information disclosure. Specifically, the bid levels for the top five ad positions in Xi'an significantly increased three days post-implementation. This disparity between the two cities persisted until the end of September, coinciding with the conclusion of the peak wedding advertising season. For mid and lower-position ad groups, no substantial difference in bid levels was observed following the bid recommendation release (see Appendix \ref{lowerplots}). This analysis suggests that the primary impact of bid recommendations is on the top ad positions, leading to increased bid levels and CPC in Xi'an.

In the next subsection, we present Difference-in-Differences regression results to quantify the impact of information disclosure to bid and CPC. 

\subsubsection{Difference-in-Differences Regression on the Impact of Information Disclosure}
\hfill\\
\indent To assess the impact of introducing bid recommendations, we utilize a difference-in-differences approach comparing changes in bids and CPCs between Xi'an and Chengdu. Chengdu was chosen as the control city due to its closely matched average bid, CPC, number of ads, and advertising revenue with Xi'an, one month prior to the policy change. Formally, we analyze the effects of bid recommendations on bids and CPC using the following regression model:

\small
\begin{center}
$Y_{it} = \alpha_0 * (T_i* Post_t) + \sum^3_{g=1}\alpha_g group_g +   \sum^3_{g=1} \beta_g * group_g*(T_i* Post_t) + fe^a_i +fe^d_t + \epsilon_{it} . $
\end{center}
\normalsize 

\indent In our regression model, the unit of observation is ad-per-day, where \( i \) denotes an ad and \( t \) a date. \( Y_{it} \) includes two outcome measures: bid and CPC. \( Post_t \) is an indicator variable for dates after the information release on August 13, 2019. \( T_i \) indicates whether ad \( i \) is in the treatment city, Xi'an.
\( group_g \) indexes ad categories by position, divided into three groups: top, middle, and low positions. Specifically, \( group_1 \) is an indicator variable set to one for ads in the top 5 positions, excluding those reserved for organic search. \( group_2 \) covers ads in positions 6 to 25, and \( group_3 \) includes ads below position 25. \( fe^a_i \) and \( fe^d_t \) control for ad-specific and day-specific effects that are constant over time, respectively. The coefficient \( \alpha_0 \) captures the difference-in-differences effect of information disclosure, while \( \beta_g \) explores the heterogeneous effects of the treatment across different ad positions. Standard errors are clustered at the ad level across all regressions to account for repeated panel data, consistent with \cite{bertrand2004much}.

\begin{table}[htb!]
\begin{center}
    {\small 
\def\sym#1{\ifmmode^{#1}\else\(^{#1}\)\fi}
\begin{tabular}{l*{4}{c}}
\hline\hline
            &\multicolumn{1}{c}{(1)}&\multicolumn{1}{c}{(2)}&\multicolumn{1}{c}{(3)}&\multicolumn{1}{c}{(4)}\\
            &\multicolumn{1}{c}{bid}&\multicolumn{1}{c}{bid}&\multicolumn{1}{c}{cpc}&\multicolumn{1}{c}{cpc}\\
\hline
T*Post         &      2.193\sym{***} (0.394) 
            &        0.633\sym{***}  (0.207)     
            &       0.669\sym{***} (0.152) 
            &       0.168\sym{*}   (0.0906)       \\
group1     &                     &       0.256    (0.243)        &                     &      -0.0695 (0.601)         \\
group2     &                     &      0.217\sym{*}  (0.130)      &                     &       0.129\sym{***} (0.0487)  \\
T*Post*group1&                     &       5.372\sym{***}(1.203) &                     &       1.902\sym{***}(0.478) \\
T*Post*group2&                        &       1.705\sym{***}(0.289) &                     &       0.436\sym{***}(0.0991)  \\
\hline
\(N\)       &       73044         &       73044         &       73044         &       73044         \\
adj. \(R^{2}\)&       0.903         &       0.908         &       0.866         &       0.868         \\
\hline\hline
\multicolumn{5}{l}{\footnotesize Standard errors in parentheses; Ad and day fixed effects, cluster at advertiser level}\\
\multicolumn{5}{l}{\footnotesize \sym{*} \(p<0.10\), \sym{**} \(p<0.05\), \sym{***} \(p<0.01\)}
\end{tabular}
}

\end{center}
\vspace{0.5em}
\caption {Difference-In-Differences Regression Table} 
\label{regtable}
\vspace{-1em}
\end{table}

To assess the impact of bid recommendation releases on market dynamics, we use daily bids averaged across bid changes within a day; if no changes occur, the average remains constant as the previous day. The difference-in-differences analysis shows an average bid increase of 2.19 RMB (0.3 USD) post-release, a 35\% rise from pre-policy levels. We further explore variations in response by ad position as detailed in column (2) of Table \ref{regtable}, which includes interactions between post-disclosure and Xi'an dummies with a group id dummy. Ads in the top five positions experience a significant 5.37 RMB (0.74 USD) increase, about twice the overall effect. In contrast, ads in positions 6 to 25 see a more modest increase of 1.71 RMB (0.24 USD). The effects are less pronounced for lower-position ads despite all receiving identical bid recommendations.

We then examine the impact on Cost Per Click (CPC), reported in column (3) of Table \ref{regtable}, which shows a 0.67 RMB (0.09 USD) increase in CPC, a 19\% lift from the pre-period level. Column (4) details the heterogeneous impact on CPC across ad groups, with top-position ads seeing the largest increase of 1.9 RMB (0.26 USD), or 28\%. Mid-position ads have a smaller, 11\% increase in CPC. These results suggest a smaller relative increase in ad prices compared to bids, which is consistent to the GSP auction mechanism where the second-highest bid sets prices.

In this section, we examine the descriptive evidence on how information disclosure affects market outcomes, specifically bids and CPC. We observe that the presence of bid-adhering bidders leads to an increase in market ad prices, potentially enhancing revenue for the ad platform. However, without known bidder valuations, the direct impact on bidder and social surplus remains unexplored. In the following section, we introduce a novel empirical model to identify heterogeneous valuations for both bid-adhering and bid-constructing bidders.

\section{Valuation Identification}
\label{sec:value_identification}

To better understand how sub-optimal bidding behaviors influence market equilibrium, specifically regarding bidder surplus and social surplus, it is crucial to identify advertisers' private values. The complexity of the GSP auction complicates empirical modeling of equilibrium bidding behavior. Additionally, the asymmetry in behaviors between bid-adhering and bid-constructing bidders adds to the challenge. To tackle these issues, we utilize distinct empirical approaches for identifying private values. For bid-constructing bidders, we characterize an equilibrium model of bidding in the GSP auction in Section \ref{sophis_model} and provide valuation identification methodology. For bid-adhering bidders, we describe the model on identifying bounds on their valuations in Section \ref{naive_model}.

\subsection{Valuation Identification of Bid-constructing Bidders}\label{sophis_model}

Here, we present a computationally feasible and practical framework for modeling the bidding strategy of bid-constructing advertisers in the GSP auction. The primary assumption for bid-constructing bidders is that they are fully rational, submitting bids that optimize their expected payoffs in equilibrium.

Specifically, we adopt a symmetric independent private valuation model within bid-constructing bidders, where bid-constructing bidders adhere to a Bayesian Nash Equilibrium bidding strategy. In our model, we analyze the daily bidding behavior of bid-constructing advertisers, indexed by \( t \). We assume that the distributions are stationary across each day. Despite advertisers having the flexibility to change bids at any time, the observed frequency of bid changes averages 4.3 times per week, aligning well with our assumption of daily modeling.

\textbf{Bidding Process :} In our daily bidding model, each bid-constructing advertiser \( i \) at period \( t \) is assumed to realize a private value per click \( v_{i,t} \), independently and identically drawn from a distribution \( V \), i.e., \( v_{i,t} \sim V \). The distribution \( V \) is our primary target for estimation, representing the valuation distribution for the bid-constructing bidders. Upon realizing their private valuation, advertiser \( i \) submits a bid \( b_{i,t} \) within the same day to maximize her expected daily payoff. The platform calculates daily quality scores for each advertiser and ranks them by the rank scores, computed as the product of the advertiser \( i \)'s submitted bid \( b_{i,t} \) and her quality score \( q_{i,t} \). The platform then orders the advertisers according to these rank scores. Let \( r_{k,t} \) be the rank score for position \( k \) at period \( t \). If advertiser \( i \) occupies the \( k^{th} \) position, her daily CTR (number of clicks) is \( \alpha_{k} \), and the realized utility is \( \alpha_k (v_{i,t} - \frac{r_{k+1,t}}{q_{i,t}}) \).

\textbf{Quality Scores and Rank Scores:} Although each consumer query generates a quality score, potentially numbering in the hundreds daily, our dataset includes only the average daily quality score for each bidder. In our model, each bid-constructing advertiser \(i\) believes her daily quality scores \(q_{i,t}\) at period \(t\) are independently and identically drawn from a distribution with cumulative distribution function (c.d.f) \(Q_{i}(q)\), which has continuous positive support on \([0, \infty)\). We assume that the distributions of quality scores and bidder valuations are independent. The rank score \(r_{k}\) for the \(k^{th}\) position is the \(k^{th}\) order statistic from the sorted product of quality scores and bids. We denote \(g_{k}(\cdot)\) as the probability density function of \(r_{k}\) and \(g_{k-1,k}(\cdot,\cdot)\) as the joint density of \(r_{k-1}\) and \(r_{k}\).

\textbf{Bidder Beliefs :} We assume that the beliefs of bid-constructing bidders are consistent with actual distributions in equilibrium, aligning with the assumptions of a Bayesian Nash Equilibrium. Bid-constructing advertisers develop beliefs about the quality score distributions, rank score distributions, and CTRs, without access to the actual realizations of these metrics or others' private values. During the identification process, we use empirical data on quality score distributions, rank score distributions, and CTRs as proxies for the formed beliefs of advertiser \( i \). 

\textbf{Estimation Process:} Each day, the bid-constructing bidder \(i\) submits a bid that maximizes her daily expected payoff, based on her realized valuation \(v_{i,t}\). The expected payoff function for bidder $i$ at time $t$ with valuation \(v_{i,t}\), $\pi(b|v_{i,t})$ is thus given by:
\begin{eqnarray}
\label{eq_profit}
\pi(b|v_{i,t}) =  \int \left[ \sum_{k=1}^{K}\alpha_{k}\int \int \mathbf{I}\left( r_{k}< {q} b <r_{k-1}\right) \left(
v_{i,t}-\frac{r_k}{{q}}\right) g_{k-1,k}\left( r_{k-1},r_{k}\right) dr_{k} dr_{k-1} \right] dQ_i(q),
\end{eqnarray}%
where $\mathbf{I}(r_{k} < qb < r_{k-1})$ is an indicator function that equals 1 if the condition $r_{k} < qb < r_{k-1}$ is satisfied, and 0 otherwise. A bid $b_{i,t}$ that maximizes the expected payoff $\pi(b|v_{i,t})$ should satisfies the following first-order condition, which is equivalent as $\frac{\pi(b|v_{i,t})}{d\ b_{i,t}}=0$:
\begin{eqnarray} 
v_{i,t}=b_{i,t}+\frac{\int \left[\sum_{k}\alpha_{k}g_{k-1}\left( {q} b_{i,t}\right) E\left(
{q}b_{i,t}-r_{k}|r_{k}<{q}b_{i,t}\right)\right]dQ_i(q) }{\int \left[\sum_{k} q g_{k}\left( {q}b_{i,t}\right) \left( \alpha_{k}-
\alpha_{k+1} \right) \right]dQ_i(q) }.
\label{gpv-like}
\end{eqnarray}
Details on deriving Equation (\ref{gpv-like}) are provided in Online Appendix \ref{app:estimation_process}. Equation (\ref{gpv-like}) employs a GPV-like formulation as outlined in \cite{guerre2000optimal}. It models the advertiser's private value \(v_{i,t}\) as a function of their equilibrium daily bid \(b_{i,t}\), the number of clicks \(\alpha_k\) at each position, and the density functions of the \(k-1^{th}\) and \(k^{th}\) position rank scores, \(g_{k-1}(\cdot)\) and \(g_{k}(\cdot)\), respectively, along with the distribution of the daily quality score \(Q_i(\cdot)\). The identification of estimation on verifying the uniqueness of bid-value mapping is provided in Online Appendix \ref{app:identification}. Notably, as both the numerator and denominator of the markdown term are positive, the resultant markup is also positive, indicating that advertisers submit bids below their actual value in equilibrium. This finding aligns with the theoretical predictions of shading behaviors in GSP (\cite{edelman2007internet}).

The basic concept of value identification is then straightforward. With the density $g_{k-1}$ and $g_{k}$, we can estimate bid-constructing bidders values from the Equation (\ref{gpv-like}). The density $g_{k-1}$ and $g_{k}$ can be approximated non-parametrically from the observed bid data. We assume that the observed pre-intervention data is generated from bidders' equilibrium bidding strategies. We employ \cite{guerre2000optimal}'s two-step estimator approach. First, we construct a sample of pseudo values from Equation (\ref{gpv-like}) using non-parametric recommendations of the density functions for the observed bids. We then use the pseudo values generated in the previous step to estimate the density of the valuation density.

To clarify, we first simulate 100 auctions daily (30 days in total) for bid-constructing bidders. In each auction, we simulate a daily quality score \(q_{i,t}\) for advertiser \(i\) from a normal distribution \( Q_i\sim N(\bar{q}_i, \epsilon)\), where \(\bar{q}_i\) represents the mean of observed quality scores for advertiser \(i\), averaged over all her observations. The standard deviation \(\epsilon\), consistent across all advertisers, is derived from the observed residual across all advertisers. This uniform variance assumption is based on the fact that the platform applies the same machine learning algorithmic treatment to compute the quality score for each advertiser, thereby standardizing the variance while allowing individual mean differences. In Online Appendix \ref{Hist_ep}, we present histogram of quality scores, supporting that normal distribution is a good approximation for quality score distribution.

We then calculate each advertiser $i$'s rank score at auction $t$, $r_{i,t} = b_{i,t} * q_{i,t}$, where $q_{i,t}$ are simulated from our estimated quality score distributions. After obtaining the auction-level empirical distribution at auction $t$, we sort the observations and obtain samples of rank scores. We then estimate the density for the 1st to 24th ad position, $g_k$, using kernel density estimation. The estimated density is given by: $g_k(x) = \frac{1}{N_k h_g} \sum_{t=1}^{N_k} K_g\left( \frac{x - r_{k,t}}{h_g} \right)$,
where $r_{k,t}$ is the $t$-th sample of the $k$-th rank score, $N_k$ is the number of observations for the $k$-th rank score, and $h_g = 1.06\sigma_k N_k^{-1/5}$ is the bandwidth, with $\sigma_k$ being the standard deviation of the samples for the $k$-th rank score. The function $K_g(\cdot)$ represents the normal kernel.

Using Equation~(\ref{gpv-like}), we estimate the pseudo values and compute the private value density, where the bid-constructing bidder's valuation distribution $V$ has density $f(x)$:
\begin{equation}
\label{f_v}
f(x) = \frac{1}{N_f h_f} \sum_{i=1}^{N_f} K\left(\frac{x - v_i}{h_f}\right).
\end{equation}
Here, $v_i$ represents the $i$-th pseudo value sample of bid-constructing bidders from Equation~(\ref{gpv-like}), $N_f$ is the number of observations, and $h_f = 1.06 \sigma_f N_f^{-1/5}$ is the bandwidth, with $\sigma_f$ denoting the standard deviation of the bid samples. 

We use pre-information disclosure data to estimate the valuations of bid-constructing bidders, isolating the effect of information disclosure. This is crucial as information disclosure influences the strategy of bid-adhering bidders and, consequently, the equilibrium rank scores. And the equilibrium rank score distribution becomes a function of bid recommendations, complicating the estimation process.

We define bid-constructing bidders as those who have bid higher than the lowest bid recommendation at least once but have never adhered to any bid recommendations. Before the platform released its bid recommendations, we observed \(1,090\) out of \(8,151\) total daily bid observations for \(62\) bid-constructing bidders. Firstly, we estimate $g_k$. Then we compute pseudo valuation, using Equation (\ref{gpv-like}), applying it only to bids from bid-constructing bidders that meet the condition \(\mathbf{I}(r_k < qb < r_{k-1})\), as in Equation (\ref{eq_profit}). Using this pseudo data, we then estimate the private values density function with Equation (\ref{f_v}). The empirical estimation on bid-constructing bidder valuation distribution and corresponding bid distribution are provided in Online Appendix \ref{app:estimation_constructing}.

\subsection{Valuation Bounds for Bid-adhering Bidders} \label{naive_model}
 
In Section \ref{sec:descriptive}, we observed the sub-optimal behaviors of bid-adhering bidders, indicating that the equilibrium bidding strategies previously discussed might not be applicable for this group. While it is possible to develop a specific stylized model for bid-adhering bidders, the primary focus of this paper is to document such behaviors and propose policy interventions, rather than to elaborate on the underlying behavioral models. Therefore, the comprehensive exploration of reasons why some bidders adhere to bid recommendations and the creation of specific models for bid-adhering bidders are left for future research, possibly involving laboratory experiments, with preliminary explanations in Section \ref{subsec:descriptive_explain}. 

Given this focus, instead of developing specific stylized models, we propose an incomplete model based on two foundational assumptions: bidders never bid more than their valuations (\cite{li2017obviously}), and they do not allow an opponent to win at a price they are willing to match. 
We assume that bid-adhering bidders share the same valuation distribution. Although we cannot precisely estimate this distribution, two foundational assumptions enable us to establish bounds on their valuations. The first assumption on no overbidding provides lower bounds on bid-adhering bidders' private valuations, as their submitted bids. The second assumption suggests that whenever a bid-adhering bidder submits a bid that is lower than one specific recommended bid, it implies that the bidder has a private valuation lower that recommendation. Otherwise, knowing that some bid-adhering bidder would bid at the recommended price that they are willing to match, the bidder would have bid at that price instead. Hence, bid recommendations that exceed submitted bids act as upper bounds for the valuation of bid-adhering bidders.

Specifically, when a bid-adhering bidder submit a bid $b_{i,t}$, the lower bound of valuation $\underline{v}_{i,t}=b_{i,t}$, and the upper bound $\overline{v}_{i,t}$ equals to the smallest bid recommendation above $b_{i,t}$, $\overline{v}_{i,t}=\min_{k}\{s_{i,k,t}, \ s_{i,k,t}>b_{i,t}\}$. We further non-parametrically estimate the lower bounds and upper bounds from $\underline{v}_{i,t}$ and $\overline{v}_{i,t}$. The empirical estimation on bid-constructing bidder valuation distribution is provided in Online Appendix (\ref{app:estimation_adhering}).

\section{Design Information Disclosure : Counterfactual Analysis}

\subsection{Preliminaries}

In Section \ref{sec:value_identification}, we provide the methodology for identifying valuations of heterogeneous bidders under generalized second price (GSP) auction. With these valuations identified, we are able to examine both bidder surplus and social surplus under the effects of the information disclosure policy. Furthermore, we conduct counterfactual analyses to guide enhancements in information disclosure design, allowing the platform to pursue various objectives such as revenue maximization or social/bidder surplus improvement.

To facilitate counterfactual analysis, we first assume that bid-constructing bidders are rational, and bid in best response to the opponents strategies. As there are multiple equilibria in GSP auctions, the following assumption assumes that bid-constructing bidders adopt a locally envy-free equilibrium bidding strategy (\cite{edelman2007internet}) which is in favor to the bidder. This implies that our counterfactual analysis provides a conservative estimate on platform revenue. And the platform revenue in any other locally envy-free equilibrium would be at least as high as what we have reported.

\begin{assumption}
\label{asmp:envy_free}
\textbf{\textit{Envy-Free Equilibrium Strategy}}: Bid-constructing bidders bid under a locally envy-free equilibrium strategy.
\end{assumption}

For bid-adhering bidders, the assumptions are more complex. We first establish the following assumption regarding the behavior of bid-adhering bidders when the platform varies the number of slots for providing bid recommendations while maintaining the same mechanism for generating these recommendations.

\begin{assumption}\label{asmp:adhere_unchange} \textbf{\textit{Stability of Bid-Adherence Under Fixed Mechanism}} : 
The bidding strategy of bid-adhering bidders remains unchanged, if the platform only varies the number of slots for providing bid recommendations, under the same recommendation generation algorithm.
\end{assumption}

Assumption \ref{asmp:adhere_unchange} is crucial. As outlined in Section \ref{subsec:info_disclosure_mechanism}, we documented the bid recommendation generation algorithm, which the platform fully discloses. While it may seem presumptuous to assume that bid-adhering bidders will consistently follow (with positive probability) all platform-provided bid recommendations under any bid recommendation generation algorithm, it is reasonable to expect that as long as the recommendation generation algorithms remain unchanged, the same group of bidders will continue to occasionally adhere to these recommendations. Conversely, bid-constructing bidders will continue to submit their own bids. A mild assumption like Assumption \ref{asmp:adhere_unchange} facilitates the conduct of counterfactual analysis by varying the amount of information provision, through varying the number of slots with bid recommendations. Altering the bid recommendation generation algorithm could potentially increase profits or enhance surplus, assuming that the strategies of bid-adhering bidders remain unchanged. However, as we cannot verify whether bid-adhering bidders would continue to adhere to bid recommendations under a different algorithm, we conduct our counterfactual analysis under Assumption \ref{asmp:adhere_unchange}, maintaining the current recommendation generation algorithm.

With Assumption \ref{asmp:adhere_unchange}, we further explore different bidding models of bid-adhering bidders. As discussed in Section \ref{subsec:descriptive_explain}, the overall bid-adherence rate among the $88$ identified bid-adhering individuals is $13.5\%$. Among those actively participating in auctions with more than 10 bid-change event, the bid-adherence rate is $12.1\%$. With the descriptive data on bid-adhering bidders, we assume a probabilistic bid-adhering strategy: when a bid-adhering bidder observes a suggestive bid, and if his valuation exceeds any of the suggestive bids, the bidder follows the suggestive bid closest to his valuation with a fixed probability.

\begin{assumption}\label{asmp:adhere_prob} 
\textbf{\textit{Probabilistic Bid-Adherence}} : When observing a set of bid recommendations $\{s_{i,k,t}\}$, a bid-adhering bidder $i$ with valuation $v_i$ adheres to the bid recommendation $s_{i,k^*(i),t} = \max_{k} \{ s_{i,k,t} \mid s_{i,k,t} < v_i \}$ with a constant probability $p$.
\end{assumption}

Assumption \ref{asmp:adhere_prob} is relatively restrictive. Firstly, it assumes that all bid-adhering bidders follow the bid recommendation with a fixed probability \( p \). Adjusting the model to fit an individual probability of adherence for each bidder may result in high variance in the estimation due to the limited number of bid-changing events per bidder. Therefore, we treat bid-adhering bidders as a homogeneous group with a constant probability \( p \). Secondly, it assumes that the bid-adhering probability is an exogenous parameter, not influenced by the level of bid recommendation. This precludes the likelihood that bid-adhering bidders are biased towards following higher or lower recommendations. In our data, we did not find such biased tendency among bid-adhering bidders. Online Appendix \ref{app:support_prob} provides data on the bids submitted and the bid recommendations observed for several bid-adhering bidders, supporting our assumption of probabilistic bid-adherence. In the counterfactual analysis, we use a grid search to find the probability $p^*$  to match the overall bid-adherence rate in table \ref{table:bid_adhere_frequency} under current information provision mechanism . 

Lastly, we still need to propose bidding model on how the bid-adhering bidder bids, when they are not following bid-recommendations. We propose two models for the bid-adhering bidders. The polynomial bidding model and the random shading model. It is important to note that due to the lack of comprehensive survey data from bid-adhering bidders, and controlled laboratory experiments, we are unable to fully characterize a bidding model for bid-adhering bidders. Nonetheless, the results presented in this counterfactual analysis section remain robust across the different bidding models we have proposed.

For the polynomial bidding model, we assume that when bid-adhering bidders do not adhere to the bid recommendations, they utilize the same bidding strategy, modeled by a polynomial function of their valuation. Specifically, if bidder \(i\) values an item at \(v_{i,t}\) on day \(t\) and chooses not to adhere to the bid recommendation, bidder \(i\) submits a bid \(b_{i,t}\) calculated as: $
b_{i,t} = p(v_{i,t})$, where $p(\cdot)$ is a polynomial function. We fit the polynomial function $p(\cdot)$ following a data-driven approach, where we utilize both lower and upper bounds of bid-adhering bidder valuations (detailed in Online Appendix \ref{app:bid_adhering_strategy}). Specifically, we simulate 1001 valuations from each estimated valuation distribution bound for bid-adhering bidders, sorting the valuations to map observed bids in a monotonic increasing relationship. This process is repeated 10,000 times, with aggregation over all samples to determine the polynomial parameters. Under the polynomial bidding model, the fitted bidding functions of the bid-adhering bidders are : for the lower-bound valuation, $p(v) = 1.11 \times 10^{-5} * v^3 - 5.54 \times 10^{-3} * v^2 + 0.83* v - 1.68$, and for the upper-bound valuation, $p(v) = -1.50 \times 10^{-3} * v^2 + 0.57* v - 3.11$. 

For the random shading model, we assume that when bid-adhering bidders deviate from bid recommendations, they randomly shade their valuations by bidding \(b(v_i) = \alpha_{i,t} v_i\), where \(\alpha_{i,t}\) is drawn from a standard uniform distribution, \(\alpha_{i,t} \sim \text{Unif}[0,1]\). This model is supported by empirical evidence: the ratio of the expected bid to the expected lower-bound of valuation is \(0.5284\), and the ratio of the expected bid to the expected upper-bound of valuation is \(0.4071\). These ratios, both close to \(0.5\), to some extent, provide justification for the uniformly random shading model.

Lastly, we generate quality scores from the estimated distributions, accounting for group heterogeneity by estimating these distributions separately for bid-constructing and bid-adhering groups. We use simulated quality scores instead of mapping actual quality scores in our dataset to each bidder mainly for confidentiality reasons. We further utilize the empirical average click-through rate (CTR) to simulate the market for the top 20 slots in sponsored search (slots beyond the top 20 exhibited less than 0.1 daily CTR).  We simulate the market with 99 bid-constructing bidders and 88 bid-adhering bidders, matching our empirical data. 

\subsection{Counterfactual Analysis}

In the counterfactual analysis, we keep the same recommendation generation algorithm and explore how variations in the amount of information provided on online advertising platforms influence market outcomes. Specifically, we explore how altering the top number of slots with bid recommendations will change the ad prices, the platform profits, and the bidder and social surplus.

Table \ref{table:analysis_p} and \ref{table:analysis_r} provide the counterfactual analysis, we present a range of of market outcomes, with the range corresponding to using the lower bound valuation to the upper bound valuation for bid-adhering bidders. For each scenario, we simulate with $10,000$ periods, and average across the observations. The numbers in brackets correspond to using the upper bound of bid-adhering bidders valuation.

In terms of platform revenue, as expected, increasing the amount of information provided, under the bid-adhering mechanism, leads to an increase in ad prices and platform revenue. This is intuitive: under our probabilistic bid-adherence assumption, bid-adhering bidders now have a higher probability of accepting a recommended bid instead of shading their bids, directly affecting the rise in ad prices and revenue. Additionally, bid-constructing bidders, knowing that bid-adhering bidders are bidding at recommended prices, also increase their bids in response, which further elevates ad prices and revenue. 

We define the surplus for bidder $i$ who occupies position $k$ at period $t$ with quality score $q_{i,t}$ as \( \alpha_k (v_{i,t} - \frac{r_{k+1,t}}{q_{i,t}}) \). The increase in platform revenue comes at a cost, with a decrease in bidder surplus. This is because, when bid-adhering bidders do not adhere to recommendations, they tend to shade their bids excessively, leading to a form of semi-collusion in the generalized second price auction. Although they do not secure the most desirable positions warranted by their high valuations, they manage to obtain reasonably lower spots at significantly reduced prices. However, when they follow the bid recommendation, they secure top sponsored search positions at the cost of increased bids. This decrease in bidder surplus does not account for the indirect benefits of occupying a higher spot, such as enhanced brand awareness and increased reviews, which are consistent with our descriptive findings on reviews (provided in Online Appendix \ref{app:review}). These effects, while not directly quantifiable in bidder surplus calculations, could potentially benefit the bid-adhering bidders.

The impact on social surplus is more complex, and there exists a trade-off between increased platform revenue and reduced bidder surplus. Both the polynomial and random bidding models suggest that releasing bid recommendations for all top 20 positions could be overly aggressive, as the decrease in bidder surplus begins to outweigh the increase in platform revenue. Our counterfactual analysis suggests that releasing a moderate level of bid recommendations for the top 10 slots results in the highest social surplus.

\setlength{\tabcolsep}{6pt}
\begin{table}[htb!]
\small
\resizebox{\columnwidth}{!}{%
\begin{tabular}{|p{3cm}|p{3cm}|p{3cm}|p{3cm}|p{3cm}|}
\hline
\textbf{Number of Slots} & \textbf{Avg. Ads Price} & \textbf{Revenue} & \textbf{Bidder Surplus} & \textbf{Social Surplus} \\ \hline
Pre-intervention & $28.23\ (28.07)$ & $3.81\times10^{4}\ (3.83)$ &$3.39\times10^{4}\ (4.03)$ & $7.20\times10^{4} \ (7.86)$ \\ \hline
Top $1$ & $34.84\ (37.76)$ & $5.93\times10^{4} \ (6.41)$ &$2.87\times10^{4}\ (3.04)$ & $8.80\times10^{4}\ (9.45)$ \\ \hline
Top $3$ (current) & $38.56\ (42.48)$ &$6.75\times10^{4}\ (7.30)$ & $2.40\times10^{4} \ (2.50)$ & $9.14\times10^{4} \ (9.80)$ \\ \hline
Top $5$ & $41.45\ (46.17)$ &$7.24\times10^{4} \ (7.90)$ & $2.07\times10^{4}\ (2.12)$ & $9.31\times10^{4} \ (10.02)$ \\ \hline
Top $10$ & $46.78\ (53.16)$& $7.86\times10^{4}\ (8.66)$ &$1.62\times10^{4} \ (1.61)$ &$9.48\times10^{4} \ (10.27)$ \\ \hline
All Positions  & $51.40\ (60.58)$ & $8.10\times10^{4} \ (9.02)$ &$1.35\times10^{4}\ (1.25)$ & $9.44\times10^{4} \ (10.26)$ \\ \hline
\end{tabular}
}
\vspace{0.1 em}
\caption{Counterfactual Analysis of Varying Number of Slots on Market Outcomes\\ 
\vspace{-0.7em}
{\normalfont  \small  Polynomial Bidding Model, Ads price and Revenue in unit of RMB}}
\label{table:analysis_p}
\end{table}
\vspace{-1em}

\setlength{\tabcolsep}{6pt}
\begin{table}[htb!]
\small
\resizebox{\columnwidth}{!}{%
\begin{tabular}{|p{3cm}|p{3cm}|p{3cm}|p{3cm}|p{3cm}|}
\hline
\textbf{Number of Slots} & \textbf{Avg. Ads Price} & \textbf{Revenue} & \textbf{Bidder Surplus} & \textbf{Social Surplus} \\ \hline
Pre-intervention & $34.73\ (30.60)$ & $5.96\times10^{4}\ (6.79)$ &$2.38\times10^{4}\ (2.53)$ & $8.33\times10^{4} \ (9.32)$ \\ \hline
Top $1$ & $38.28\ (43.90)$ & $6.68\times10^{4} \ (7.51)$ &$2.33\times10^{4}\ (2.36)$ & $9.01\times10^{4}\ (9.87)$ \\ \hline
Top $3$ (current) & $41.34\ (47.64)$ &$7.22\times10^{4}\ (8.05)$ & $2.05\times10^{4} \ (2.06)$ & $9.27\times10^{4} \ (10.01)$ \\ \hline
Top $5$ & $43.70\ (50.57)$ &$7.53\times10^{4} \ (8.37)$ & $1.86\times10^{4}\ (1.84)$ & $9.39\times10^{4} \ (10.21)$ \\ \hline
Top $10$ & $47.99\ (55.80)$& $7.92\times10^{4}\ (8.79)$ &$1.55\times10^{4} \ (1.50)$ &$9.47\times10^{4} \ (10.29)$ \\ \hline
All Positions  & $51.67\ (60.84)$ & $8.09\times10^{4} \ (9.01)$ &$1.34\times10^{4}\ (1.24)$ & $9.44\times10^{4} \ (10.25)$ \\ \hline
\end{tabular}
}
\vspace{0.1 em}
\caption{Counterfactual Analysis of Varying Number of Slots on Market Outcomes\\ 
\vspace{-0.7em}
{\normalfont  \small  Random Bidding Model, Ads price and Revenue in unit of RMB}}
\vspace{-1em}
\label{table:analysis_r}
\end{table}

\vspace{-2em}
\section{Conclusion}

Identifying the heterogeneity in bidder valuations and strategies is essential for optimizing outcomes on online advertising platforms under information disclosure. This study explores the design of information disclosure policies and their impact on market dynamics within a Yelp-like platform that introduced a bid recommendation system. Our analysis identifies a subset of advertisers (termed bid-adhering advisors), who, characterized by lower review ratings and initial ad placements, tend to follow the platform's bid recommendations -- a theoretically sub-optimal practice. In contrast, bid-constructing bidders consistently craft their own bids without following released recommendations.

Empirical evidence from a difference-in-differences (DID) analysis indicates that bid recommendations elevate ad prices and platform revenue. To thoroughly assess the impact of these recommendations on participant surplus, it is necessary to identify bidder valuations within this intricate system featuring heterogeneous reaction to information disclosure. Our empirical model is the first to identify private values for bid-constructing advertisers in the online Generalized Second Price (GSP) auction, and establishes valuation bounds for bid-adhering bidders. Utilizing these valuations, we conduct counterfactual analysis to explore how varying the number of disclosed bid recommendations affects platform revenue, bidder surplus, and total surplus. Our findings suggest that while platform profits increase with increased information disclosure, social surplus peaks under a carefully designed information provision scheme, highlighting that more extensive disclosure is not always socially optimal.

This paper underscores the importance of recognizing and accounting for heterogeneous responses to information disclosure. By pinpointing a subgroup of market participants more likely to adhere to recommendations, platforms can improve their information disclosure strategies. The implications of our findings extend beyond search advertising to other online marketplaces, such as Amazon and Airbnb, which provide price recommendations to suppliers based on demand and competitive dynamics.

There are many promising directions for future research. First, behavioral economics could provide significant insights into the psychological drivers of bid adherence through controlled lab experiments, which were not covered in this paper due to resource constraints. Additionally, investigating the effects of changes in the recommendation algorithm on bidder behavior could enhance our understanding of bidder sensitivity and adaptability. Another fruitful area of study is the interaction between optimal auction design and bid-adherence behavior. For instance, while the platform currently maintains a reserve price of 2.5 RMB in the bid recommendation system, future research could explore how to optimally set reserve prices to maximize platform objectives in environments characterized by information disclosure and bid-adhering behaviors. Another interesting direction for future theory research is to formally analyze the impact of increasing the number of bid recommendation slots on revenue and social surplus with analytical models. Although quality score interventions complicate the definition of social surplus — since quality scores reflect potential consumer satisfaction and primarily benefit the consumer side—there remains an opportunity to develop theoretical models that explore the effects of information disclosure levels on ad revenue and bidder surplus without the influence of quality scores. Such an extension could provide valuable theory-guided insights into optimizing platform information disclosure strategies for advertisers.

 \newpage 
\bibliographystyle{apalike}
\bibliography{11reference}
\newpage
\section{Online Appendix}

\subsection{Proof of Theorem \ref{theorem:irrational}}\label{irrational_proof}

We first introduce the following assumptions for Theorem \ref{theorem:irrational}. 
\begin{assumption}\label{assumption:irrational} 
\begin{enumerate}
    \item Bidder $i$ is a Bayesian utility maximizer who optimizes her expected utility based on her beliefs about quality score distributions, aware of her own valuation at period $t$, denoted as $v_{i,t}$.
    \item Bidder $i$ understands the bid recommendation generation process. Specifically, for the $k^{th}$ position at period $t$, bidder $i$ receives a bid recommendation $s_{i,k,t} = \frac{r_{k+1,t-1}}{\hat{q}_{i,t}}$, where $r_{k+1,t-1}$ is the rank score of position $k+1$ from period $t-1$, and $\hat{q}_{i,t}$ is the simulated quality score for bidder $i$ at period $t$.
    \item Bidder \(i\)'s belief about the actual quality score distribution for each bidder \(j\) at period \(t\) is independently drawn from distributions \(Q_{j,t}\), which have continuous positive support on the interval \([0,  \infty)\). Similarly, the belief regarding the platform-simulated quality score distribution for calculating bidder \(j\)'s bid recommendation at period \(t\), \(\hat{q}_{j,t}\), follows distribution of \(\hat{Q}_{j,t}\) with continuous positive support on the interval \([0, \infty )\).
\end{enumerate}
\end{assumption}

With Assumption \ref{assumption:irrational}, we are ready to prove Theorem \ref{theorem:irrational}. As a Bayesian expected utility maximizer observing the quality score adjusted bid recommendation at period \( t \) for position \( k \), \( s_{i,k,t} \), bidder \( i \) updates the belief on the distribution of the \( k+1^{th} \) position's rank score from period \( t-1 \), denoted \( r_{k+1,t-1} \) which is distributed as \( s_{i,k,t} \hat{Q}_{i,t} \). Given that other bidders follow the bid recommendation, the bidding strategy for bidder \( j \) is \( b_{j} = \frac{r_{k+1,t-1}}{\hat{Q}_{j,t}} \), and the rank score associated with \( b_j \) follows the distribution \( s_{i,k,t} \frac{\hat{Q}_{i,t} Q_{j,t}}{\hat{Q}_{j,t}} \). Under Assumption \ref{assumption:irrational}, as long as the bid recommendation $s_{i,k,t}$ is non-zero, the rank score distributions at period \( t \) have continuous positive support on \( [0, \infty) \). And bidder \( i \) can update the belief on the rank score distribution $r_{k,t}$, and the joint rank score distribution with associated probability density function (pdf) \( g_{k-1,k,t}(\cdot,\cdot ) \) with support on \( [0, \infty)\times [0, \infty) \) at period $t$.

Bidder \(i\) then solves the following optimization problem to maximize her expected profit, given her valuation \(v_{i,t}\). Let \(Q_{i,t}(q)\) denote the cumulative distribution function (cdf) of the distribution \(Q_{i,t}\),
\begin{eqnarray*}
\pi(b|v_{i,t}) &=&  \int^{\infty}_{0} \left[\sum_{k}\alpha_{k}\int_{qb}^{\infty }\left( \int_{0}^{q b}\left( v_{i,t}-\frac{r_{k,t}}{{q}}\right)
g_{k-1,k,t}\left( r_{k-1,t},r_{k,t}\right) dr_{k,t} \right) dr_{k-1,t}\right] dQ_{i,t}(q)
\end{eqnarray*}%

Taking the first-order derivative and setting it to zero (for details please refer to Appendix \ref{app:estimation_process}), let $g_{k}(\cdot)$ be the marginal probability density function of the joint rank score pdf \( g_{k-1,k,t}(\cdot,\cdot) \), 
\begin{eqnarray}
v_{i,t} 
&=& b + \frac{\int^{\infty}_{0} \left[\sum_{k}\alpha_{k}g_{k-1}\left( q b\right) E\left(
qb-r_{k,t}|r_{k,t}<qb\right)\right]dQ_{i,t}(q) }{\int^{\infty}_{0} \left[\sum_{k} q g_{k}\left( {q}b\right) \left( \alpha_{k}-
\alpha_{k+1} \right) \right]dQ_{i,t}(q) }
\label{appendix:equation:irrational1}
\end{eqnarray}

Given the valuation $v_{i,t}$, the best response on bid $b$ of a Bayesian expected utility maximizer should satisfy Equation (\ref{appendix:equation:irrational1}). As $s_{i,k,t}=\frac{r_{k+1,t-1}}{\hat{q}_{i,t}}$, where $\hat{q}_{i,t}$ is a sample of bidder $i$'s simulated quality score at period $t$ from an atomless distribution $\hat{Q}_{i,t}$, $P(s_{i,k,t}=b)=0$. The bid recommendation $s_{i,k,t}$ coincides with bidder $i$'s optimal bid $b$ is a probability zero event. Hence, any strategy that involves a positive probability on adhering to bid recommendation $s_{i,k,t}$ is not a best response strategy.
\newpage 
\subsection{Descriptive Patterns on Reviews and Review Stars}
\label{app:review}
\begin{figure}[htb!]
\centering
\begin{subfigure}{.5\textwidth}
  \centering
  \includegraphics[width=.8\linewidth]{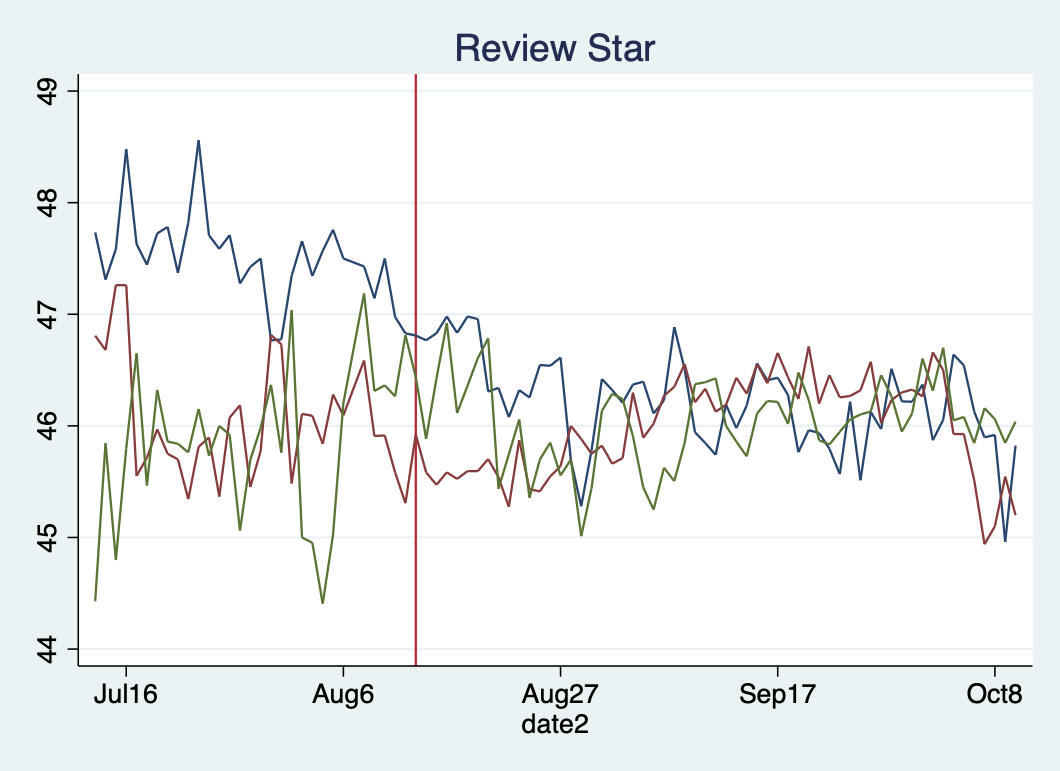}
  \label{fig:review_so}
\end{subfigure}%
\begin{subfigure}{.5\textwidth}
  \centering
  \includegraphics[width=.8\linewidth]{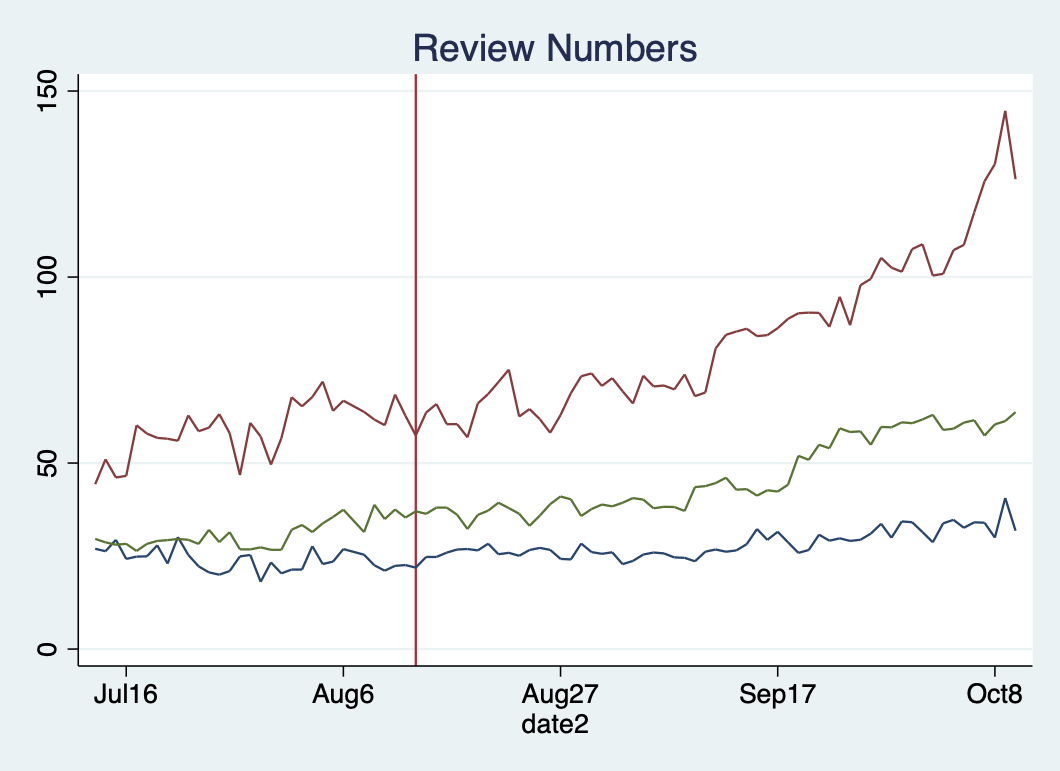}
  \label{fig:review_no}
\end{subfigure}
\caption{\centering Reviews}
\footnotesize\textsuperscript{*} Figure \ref{fig:review} displays the review star and review number for bid-constructing bidders (shown in blue) and bid-adhering bidders (shown in red). The green line represents the daily average level in the control city among top 25th-position advertisers. The vertical line reflects the intervention date.
\label{fig:review}
\end{figure}

We explore differences in reviews metrics between groups as shown in Figure \ref{fig:review}. Bid-adhering bidders have lower review stars but a higher number of reviews, suggesting they are likely to adopt a straightforward marketing strategy focused mostly on increasing transaction volumes. Following the release of bid recommendations, the review stars among the bid-adhering group have slightly increased. Additionally, as depicted in Figure \ref{fig:review}, the number of reviews for the bid-adhering group has also improved, likely due to new consumers (\cite{narayanan2015position}). In general, we observe improvements in advertising performance in terms of review numbers for bid-adhering bidders, who also incur higher costs per click after information disclosure. 

\newpage
\subsection{Pre-Post Changes in Bid and CPC in Lower Ad Positions} \label{lowerplots}

\begin{figure}[htb!]
\centering
\begin{subfigure}{.5\textwidth}
  \centering
  \includegraphics[width=.8\linewidth]{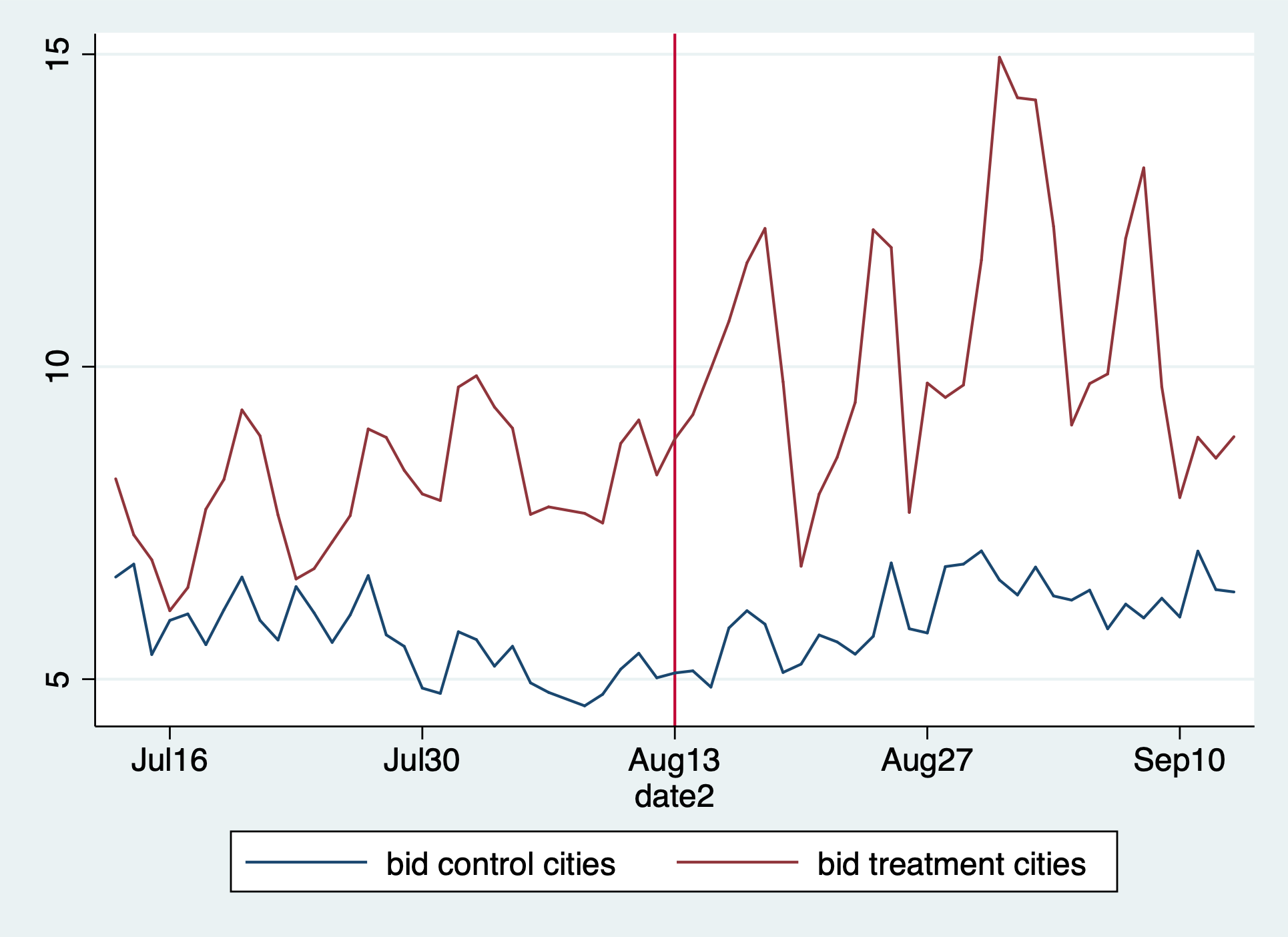}
  \captionsetup{font=footnotesize}
  \caption{Average Bid of Top 5 Ad positions}
  \footnotesize\textsuperscript{*} Figure \ref{fig:bid_app_1} displays the average daily bid level in Xi'an (red line) and Chengdu (blue line) of Mid (6-25th) Ad slots. 
  \label{fig:bid_app_1}
\end{subfigure}%
\begin{subfigure}{.5\textwidth}
  \centering
  \includegraphics[width=.8\linewidth]{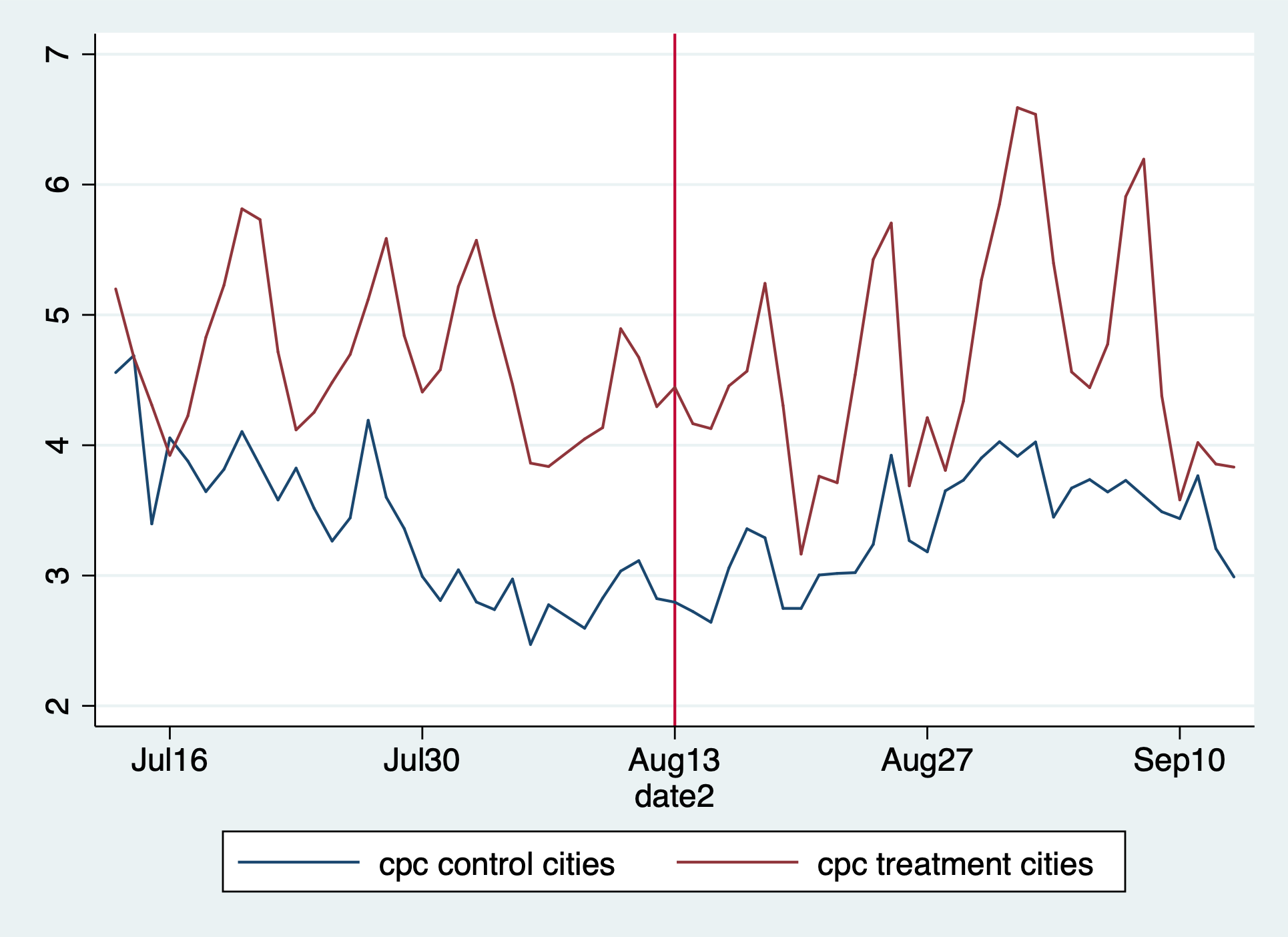}
  \captionsetup{font=footnotesize}
  \caption{Average CPC of Top 5 Ad positions}
  \footnotesize\textsuperscript{*} Figure \ref{fig:cpc_app_1} displays the average daily bid level in Xi'an (red line) and Chengdu (blue line) of Mid (6-25th) Ad slots. 
  \label{fig:cpc_app_1}
\end{subfigure}
\caption{\centering Pre-and-Post Changes in Bid and CPC of Mid (6-25th) Ad Slots}
\label{fig:market outcome_mid}
\end{figure}

\begin{figure}[htb!]
\centering
\begin{subfigure}{.5\textwidth}
  \centering
  \includegraphics[width=.8\linewidth]{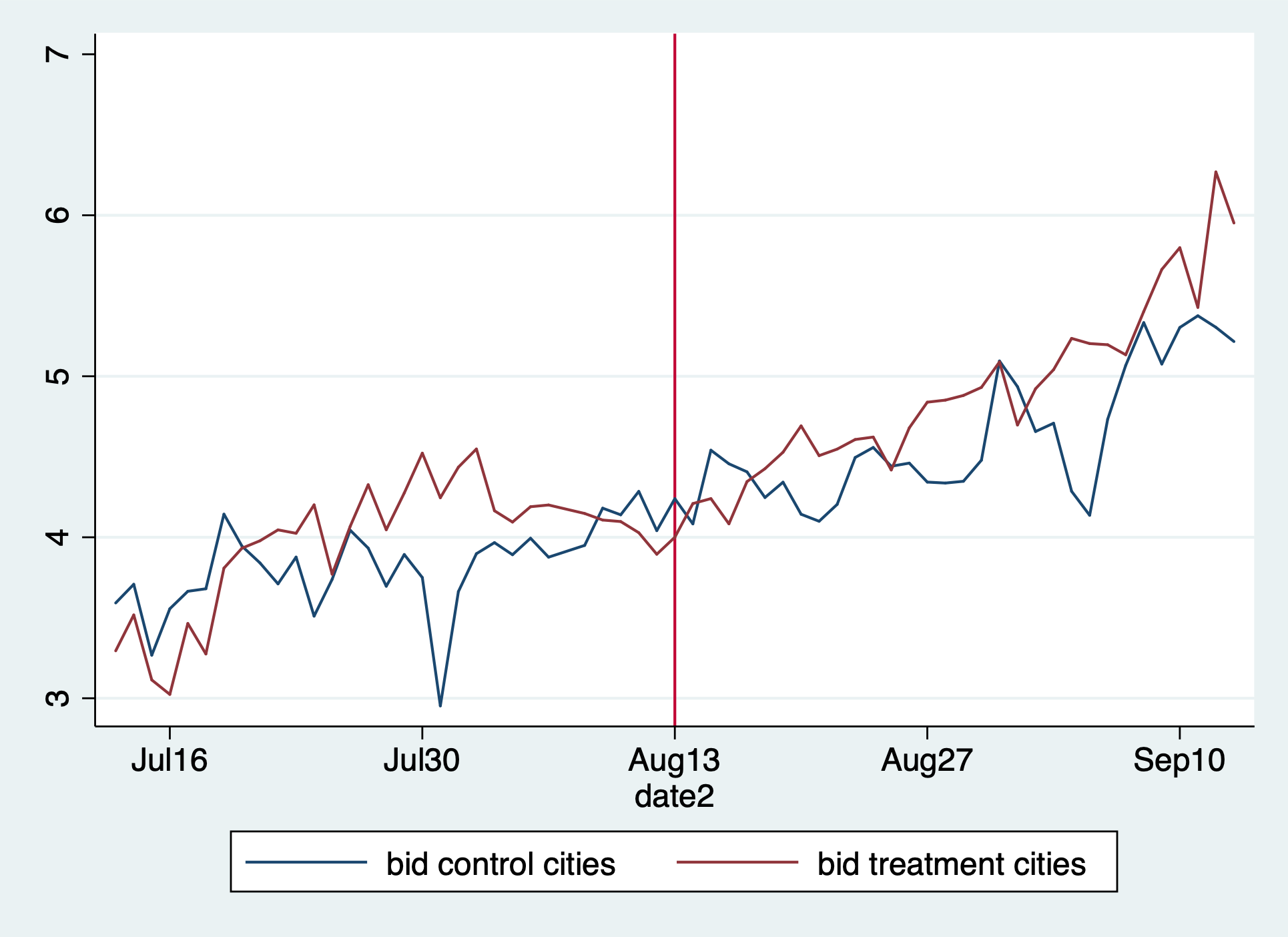}
  \captionsetup{font=footnotesize}
  \caption{Average Bid of Top 5 Ad positions}
  \footnotesize\textsuperscript{*} Figure \ref{fig:bid_app} displays the average daily bid level in Xi'an (red line) and Chengdu (blue line)  of lower ($>$25th) Ad slots.
  \label{fig:bid_app}
\end{subfigure}%
\begin{subfigure}{.5\textwidth}
  \centering
  \includegraphics[width=.8\linewidth]{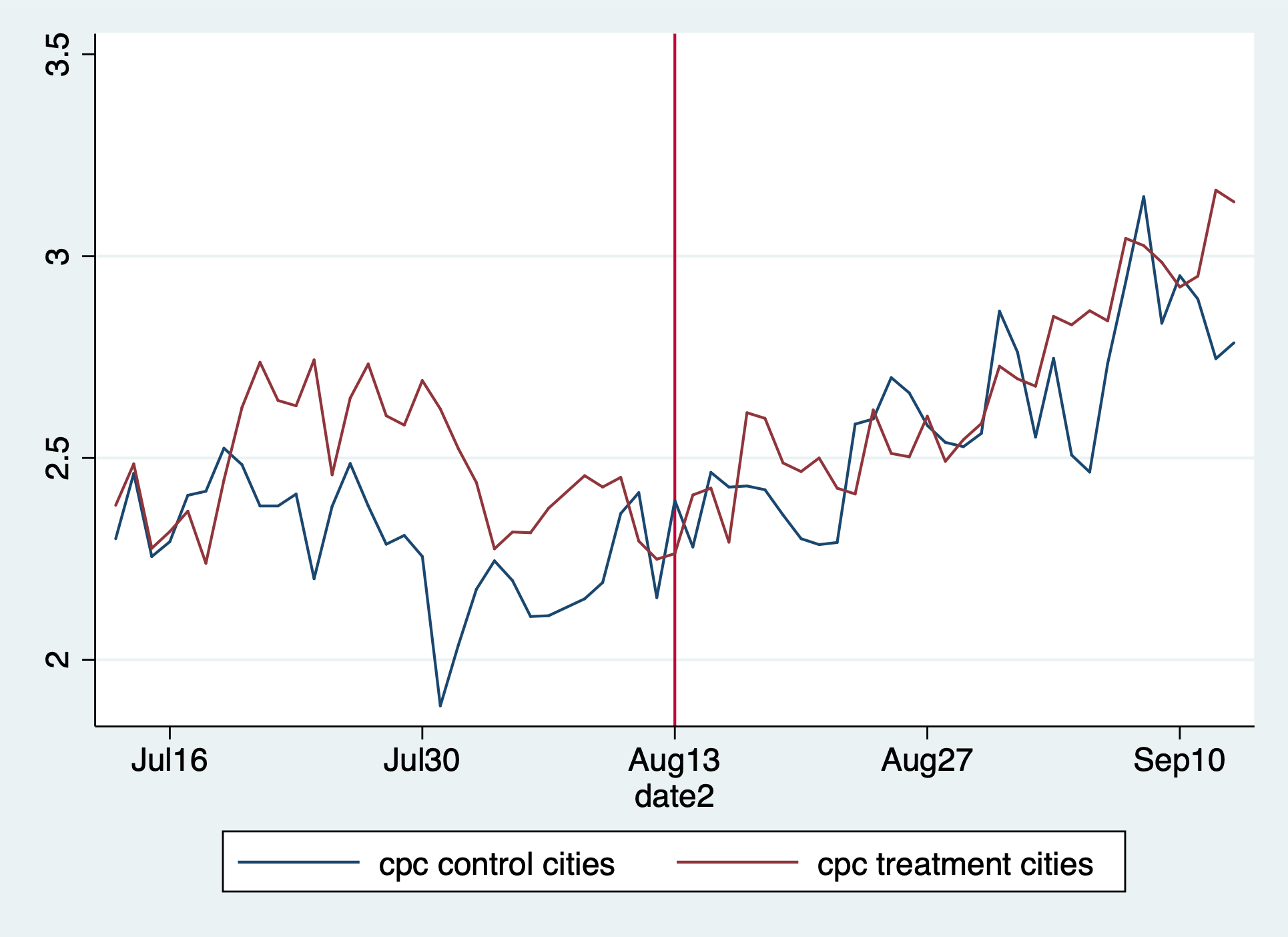}
  \captionsetup{font=footnotesize}
    \caption{Average CPC of Top 5 Ad positions}
  \footnotesize\textsuperscript{*} Figure \ref{fig:cpc_app} displays the average daily bid level in Xi'an (red line) and Chengdu (blue line) of of lower ($>$25th) Ad slots.
  \label{fig:cpc_app}
\end{subfigure}
\caption{\centering Pre-and-Post Changes in Bid and CPC of Lower ($>$25th) Ad Slots}
\label{fig:market outcome_low}
\end{figure}

\newpage
\subsection{Estimation Process}
\label{app:estimation_process}

The bidder $i$'s expected payoff when submitting bid equals to $b$, with valuation $v_{i,t}$, is given by
\begin{eqnarray*}
\pi(b|v_{i,t}) &=&\int \left[ \sum_{k=1}^{K}\alpha_{k}\int \int 1\left( r_{k}< {q} b <r_{k-1}\right) \left(
v_{i,t}-\frac{r_k}{{q}}\right) g_{k-1,k}\left( r_{k-1},r_{k}\right) dr_{k} dr_{k-1} \right] dQ_i(q)\\
&=&  \int \left[\sum_{k}\alpha_{k}\int_{{q}b}^{\infty }\left( \int_{0}^{{q} b}\left( v_{i,t}-\frac{r_{k}}{{q}}\right)
g_{k-1,k}\left( r_{k-1},r_{k}\right) dr_{k} \right) dr_{k-1}\right] dQ_i(q).
\end{eqnarray*}%

Taking the first order derivative with respect to $b$ gives us:%
\small
\begin{eqnarray*}
\frac{d \pi(b|v_{i,t})}{d b} &=& - \int \left[\sum_{k}\alpha_{k}\left( \int_{0}^{{q} b }q\left( v_{i,t}-\frac{r_{k}}{{q}}\right)
g_{k-1,k}\left({q}b,r_{k}\right) dr_{k}-q\left( v_{i,t}-b\right)
\int_{{q}b}^{\infty}g_{k-1,k}\left( r_{k-1},{q} b\right) dr_{k-1}\right)\right]dQ_i(q)  \\
&=&- \int \left[\sum_{k}\alpha_{k}\left( g_{k-1}\left({q} b\right) \int_{0}^{{q} b }q\left(
v_{i,t}-\frac{r_k}{{q}}\right) g_{k-1,k}\left( r_{k}|{q}b\right) dr_{k}-{q}\left( v_{i,t}-b\right) \int_{{q}b}^{\infty}g_{k-1,k}\left( r_{k-1},{q}b\right)
dr_{k-1}\right) \right]dQ_i(q) \\
&=&- \int \left[\sum_{k}\alpha_{k}\left( g_{k-1}\left( {q}b\right) q\left( v_{i,t}-\frac{1}{q}E\left(
r_{k}|r_{k}<{q} b\right) \right) -{q}\left( v_{i,t} -b\right) g_{k}\left( {q} b\right)
\right) \right]dQ_i(q).
\end{eqnarray*}%
\normalsize

Solving for $v$ with the first order condition $\pi^{\prime}(b|v_{i,t})=0$, we have
\begin{eqnarray*}  
v_{i,t} &=& \frac{\int \left[\sum_{k}\alpha_{k}\left( g_{k-1}\left( {q} b\right) E\left(
r_{k}|r_{k}<{q}b\right) - {q}b g_{k}\left( {q} b\right) \right)\right]dQ_i(q) }{\int 
\left[ \sum_{k}q\alpha_{k}\left( g_{k-1}\left( {q}b\right) -g_{k}\left( {q}b\right) \right)\right] dQ_i(q) } \\
&=& b + \frac{\int \left[\sum_{k}\alpha_{k}g_{k-1}\left( {q} b\right) E\left(
r_{k}-{q}b|r_{k}<{q} b\right) \right]dQ_i(q)}{\int \left[\sum_{k}q\alpha_{k}\left( g_{k-1}\left( {q} b\right)
-g_{k}\left( {q}b\right) \right) \right]dQ_i(q)} .
\end{eqnarray*}

Rearrange the above equation and let \( b = b_{i,t} \), which represents the bid submitted by bidder \( i \) at period \( t \), 
\begin{eqnarray*} 
v_{i,t}=b_{i,t}+\frac{\int \left[\sum_{k}\alpha_{k}g_{k-1}\left( {q} b_{i,t}\right) E\left(
{q}b_{i,t}-r_{k}|r_{k}<{q}b_{i,t}\right)\right]dQ_i(q) }{\int \left[\sum_{k} q g_{k}\left( {q}b_{i,t}\right) \left( \alpha_{k}-
\alpha_{k+1} \right) \right]dQ_i(q), }
\end{eqnarray*}
which is Equation (\ref{gpv-like}).

We further show that any bid maximizing the expected payoff function must satisfy Equation (\ref{gpv-like}), establishing that the first-order condition is necessary. In Appendix \ref{app:identification}, we also confirm that this condition is sufficient. For the necessity of the first-order condition, we show that a bid \(b\) that optimizes the expected payoff \(\pi(b|v_{i,t})\) must lie within the interior of \([0,\infty)\). The feasible bid set is \([0,\infty)\). And \(\pi(b=0|v_{i,t})=0\). Given that the valuation \(v_{i,t}\) is bounded, overbidding beyond this valuation is also not optimal (\cite{edelman2007internet}). Therefore, the optimal bid $b\in(0,v_{i,t})$ is within the interior of feasible bid set. Hence, the first-order condition outlined in Equation (\ref{gpv-like}) is a necessary condition for $b$ being an optimal bid that maximizes the expected payoff.

\subsection{Identification for Estimation}
\label{app:identification}

Firstly, we propose the following lemma on showing that the bidding strategy is increasing in valuation -- with a higher valuation, the optimal bid derived in Equation (\ref{gpv-like}) is higher.

\begin{lemma}
\label{lemma:increasing}
Let $b(v_{i,t})$ be the bidding strategy that maximizes the expected payoff function $\pi(b|v_{i,t})$. $b(v_{i,t})$ is monotone increasing in $v_{i,t}$.
\end{lemma}

We prove this by showing that the expected payoff of bidding $b$ given $v_{i,t}$, $\pi(b|v_{i,t})$ has an increasing difference.
\begin{eqnarray*}
\frac{\partial \pi(b|v_{i,t})}{\partial b} &=&- \int \left[\sum_{k}\alpha_{k}\left( g_{k-1}\left( {q}b\right) q\left( v_{i,t}-\frac{1}{q}E\left(
r_{k}|r_{k}<{q} b\right) \right) -{q}\left( v_{i,t}-b\right) g_{k}\left( {q} b\right)
\right) \right]dQ_i(q)
\end{eqnarray*}%
And 
\begin{eqnarray*}
\frac{\partial^2 \pi(b|v_{i,t}) }{\partial bd v_{i,t}} = \int\sum_k\alpha_k q (g_k(qb)-g_{k-1}(qb))dQ_i(q)= \int\sum_k(\alpha_{k}-\alpha_{k+1})g_{k}(qb)dQ_i(q) > 0
\end{eqnarray*}%
Hence $\pi(b|v_{i,t})$ has increasing difference in $(b,v_{i,t})$, and $b(v_{i,t})$ is monotone increasing in $v$ (\cite{milgrom1994monotone}). 

We then demonstrate that all pairs \((b_{i,t}, v_{i,t})\) satisfying Equation (\ref{gpv-like}) imply that \(b_{i,t}\) maximizes the function \(\pi(b|v_{i,t})\). Consider a pair \((b_{i,t}, v_{i,t})\) that meets the first-order condition. Let \(\underline{b} < b_{i,t} < \bar{b}\), and \((\underline{v}, \underline{b})\) and \((\bar{v}, \bar{b})\) are pairs such that \(\underline{b}\) and \(\bar{b}\) maximize \(\pi(b|\underline{v})\) and \(\pi(b|\bar{v})\) respectively, each satisfying the first-order condition under the same rank score and quality score distributions.  From Lemma \ref{lemma:increasing}, \(\underline{v} < \bar{v}\). 

Moreover, from Appendix \ref{app:estimation_process}, we have
\small
$$\frac{d \pi(b|v_{i,t})}{d b}\bigg|_{b=\underline{b}}=\left(\int \left[\sum_{k} q g_{k}\left( {q}\underline{b}\right) \left( \alpha_{k}-
\alpha_{k+1} \right) \right]dQ_i(q)\right) \left(-v_{i,t} +  \underline{b}+\frac{\int \left[\sum_{k}\alpha_{k}g_{k-1}\left( {q}\underline{b}\right) E\left(
{q}\underline{b}-r_{k}|r_{k}<{q}\underline{b}\right)\right]dQ_i(q) }{\int \left[\sum_{k} q g_{k}\left( {q}\underline{b}\right) \left( \alpha_{k}-
\alpha_{k+1} \right) \right]dQ_i(q)} \right).$$
\normalsize
And from Equation (\ref{gpv-like}), as $(\underline{v},\underline{b})$ satisfies the first order condition,
$$\underline{v} = \underline{b}+\frac{\int \left[\sum_{k}\alpha_{k}g_{k-1}\left( {q}\underline{b}\right) E\left(
{q}\underline{b}-r_{k}|r_{k}<{q}\underline{b}\right)\right]dQ_i(q) }{\int \left[\sum_{k} q g_{k}\left( {q}\underline{b}\right) \left( \alpha_{k}-
\alpha_{k+1} \right) \right]dQ_i(q)}.$$

\normalsize Hence
$$\frac{d \pi(b|v_{i,t})}{d b}\bigg|_{b=\underline{b}}=\left(\int \left[\sum_{k} q g_{k}\left( {q}\underline{b}\right) \left( \alpha_{k}-
\alpha_{k+1} \right) \right]dQ_i(q)\right) \left(\underline{v} -v_{i,t}  \right)<0$$
\normalsize
Similarly,
$$\frac{d \pi(b|v_i)}{d b}\bigg|_{b=\overline{b}}=\left(\int \left[\sum_{k} q g_{k}\left( {q}\overline{b}\right) \left( \alpha_{k}-
\alpha_{k+1} \right) \right]dQ_i(q)\right) \left(\overline{v} -v_{i,t}  \right)>0$$
Therefore, \(\frac{d \pi(b|v_{i,t})}{d b}\bigg|_{b=\underline{b}} < 0<\frac{d \pi(b|v_{i,t})}{d b}\bigg|_{b=\bar{b}}\) for \(\underline{b} < b_i < \bar{b}\). Thus, fixing valuation \(v_{i,t}\), let \(\pi'(b|v_{i,t})=\frac{d \pi(b|v_{i,t})}{d b} \). As $\pi'(b|v_{i,t})$ is a continuous function in \(b\), and when  $b^*$ satisfies $\pi'(b^*|v_{i,t})=0$, for all $\underline{b}<b^*<\overline{b}$, $\pi'(\underline{b}|v_{i,t})<0$, $\pi'(\overline{v}|v_{i,t})>0$. The derivative \(\pi'(b|v_{i,t})\) exhibits a single crossing property to $x$-axis in \((b,\pi'(b))\) plane. 

Given that the function \(\pi'(b|v_{i,t})\) is continuous and crosses the x-axis only once, the first-order condition is sufficient. This implies, the bid \(b_{i,t}\) which satisfies the first order condition described in Equation (\ref{gpv-like}) is the unique maximizer of the expected payoff function \(\pi(b|v_{i,t})\).

\medskip

\subsection{Estimation Details of Quality Score Distributions}\label{Hist_ep}

\begin{figure}[htb!] 
    \centering
	\includegraphics[width=.5\textwidth]{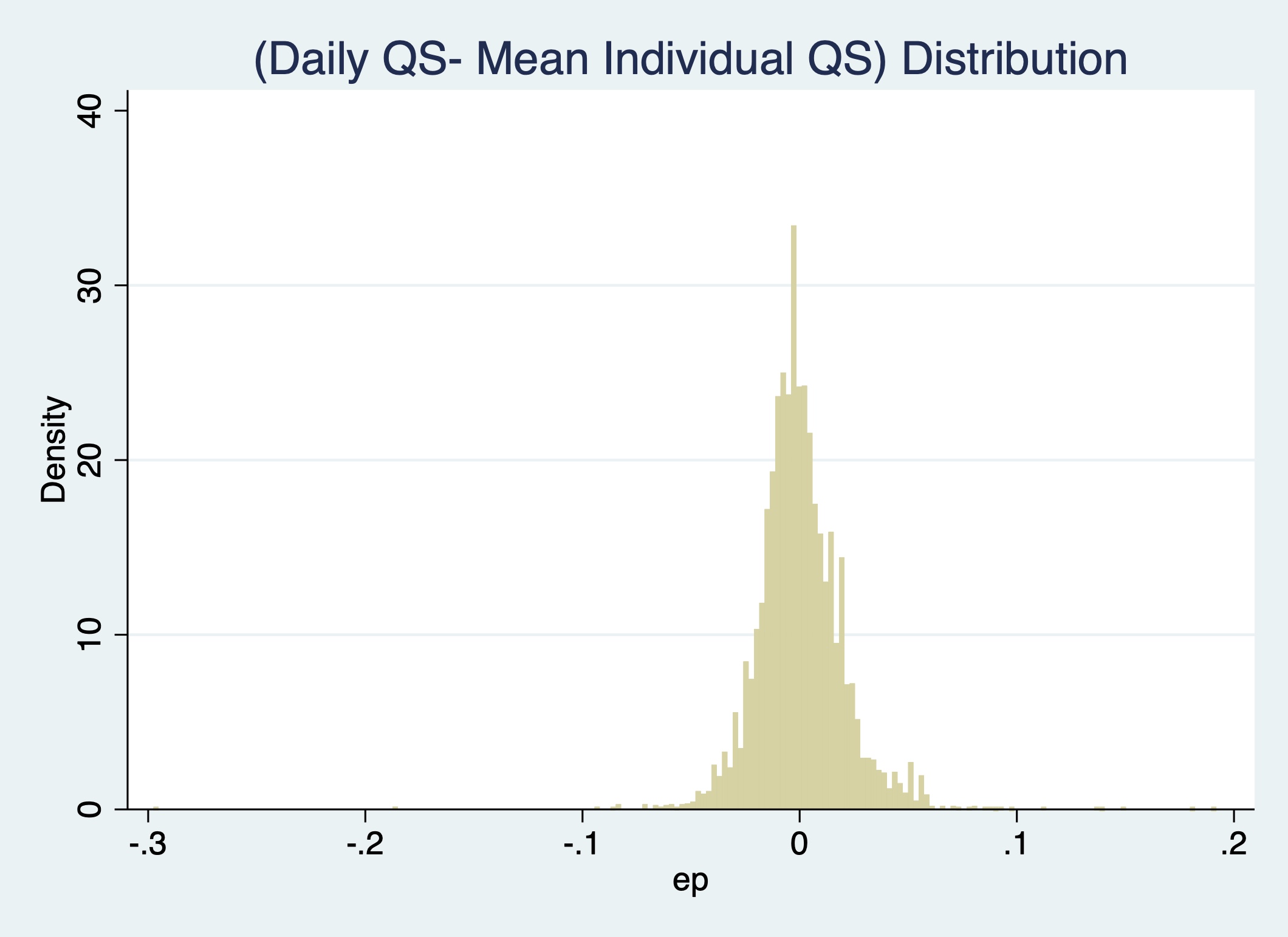}
	\caption{\centering Histogram of Daily Quality Score Variations}
	\label{fig:ep_dist}
\end{figure}

Figure \ref{fig:ep_dist} shows the histogram of the mean-centered individual quality scores, demonstrating that a normal approximation for these scores is appropriate.

\newpage 
\subsection{Empirical Bid Distribution and Estimated Valuation Distribution for Bid-constructing Bidders}
\label{app:estimation_constructing}

\begin{figure}[htb!] 
    \centering
	\includegraphics[width=.5\textwidth]{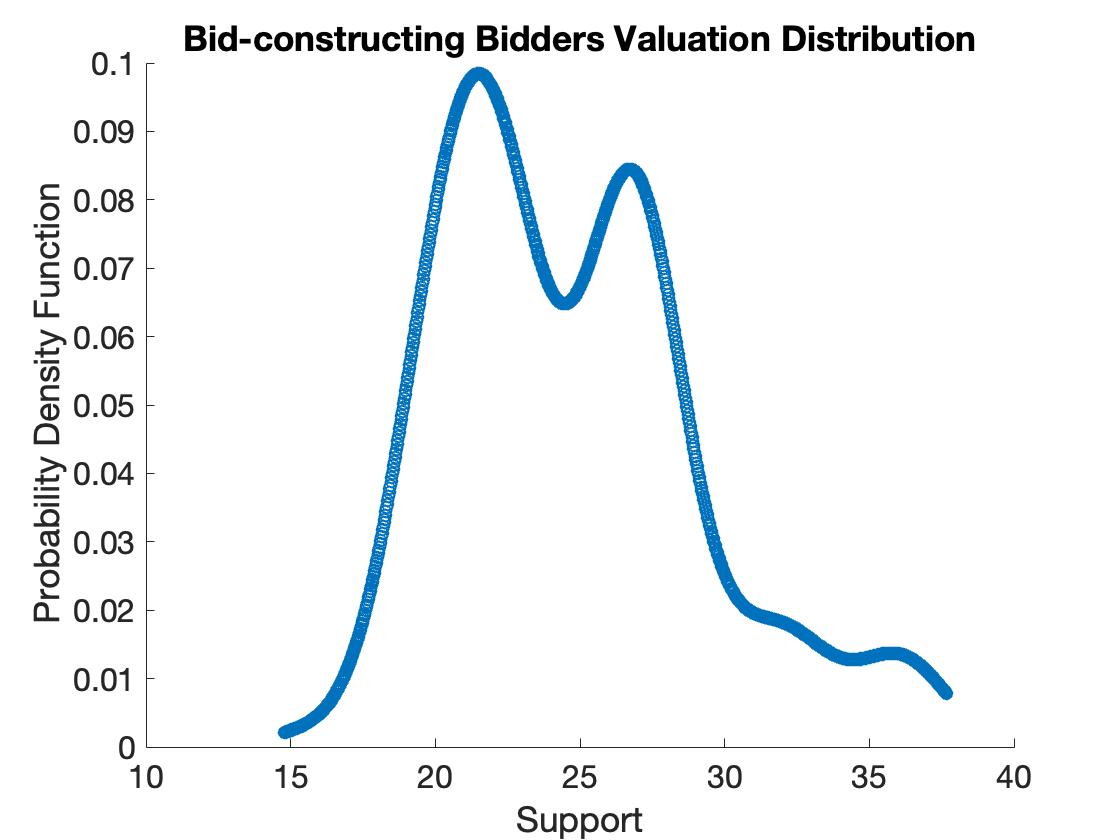}
	\caption{\centering Valuation Density of Bid-constructing Bidders}
	\label{fig:rational_density}
\end{figure}

\begin{figure}[htb!] 
    \centering
	\includegraphics[width=.5\textwidth]{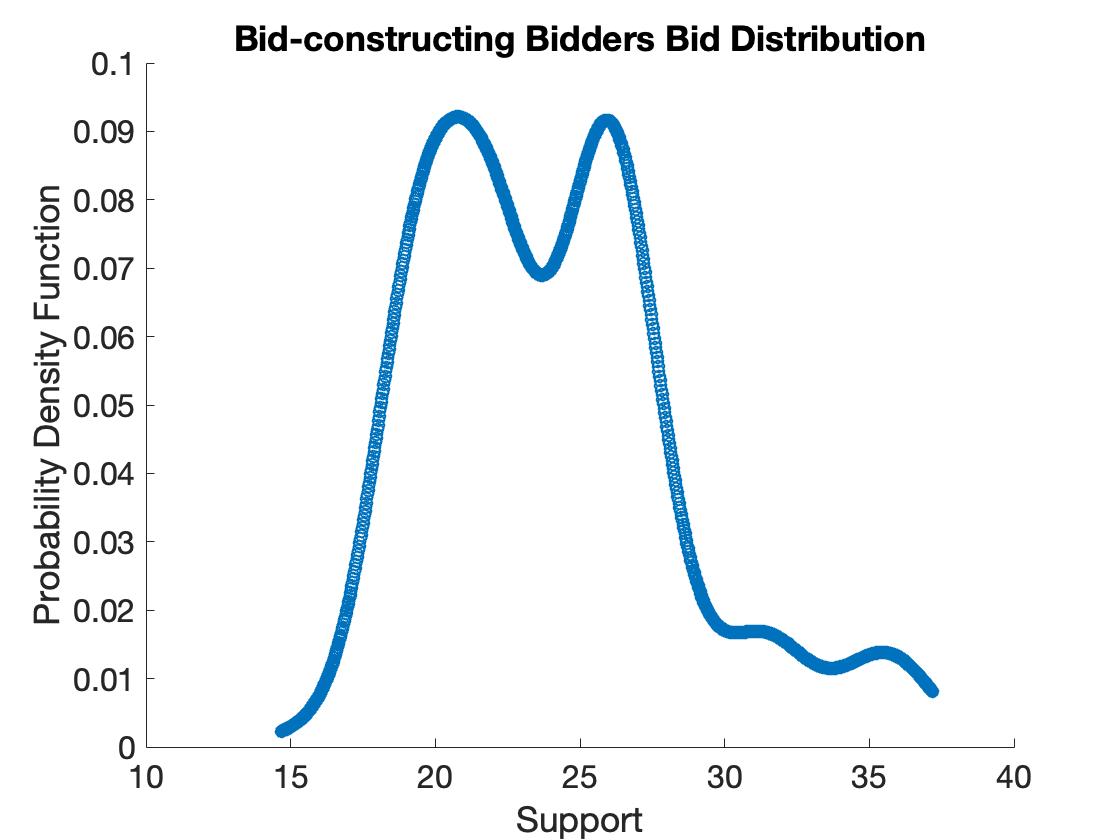}
	\caption{\centering Bid-constructing Bidders's Empirical Bid Density}
	\label{fig:bid_dist}
\end{figure}

Figure \ref{fig:rational_density} displays the estimated value density for the bid-constructing bidders. The estimated valuation distribution approximates a bimodal distribution. This aligns with the bimodal bid distribution, detailed in Figure \ref{fig:bid_dist}. The bimodal distribution suggests that there might be further segmentation of bidders within the bid-constructing bidders (e.g. high types and low types), a phenomenon similarly documented in \cite{hernandez2020estimation}.

\newpage
\subsection{Estimated Valuation Bounds for Bid-adhering Bidders}
\label{app:estimation_adhering}

\begin{figure}[htb!] 
    \centering
\includegraphics[width=.5\textwidth]{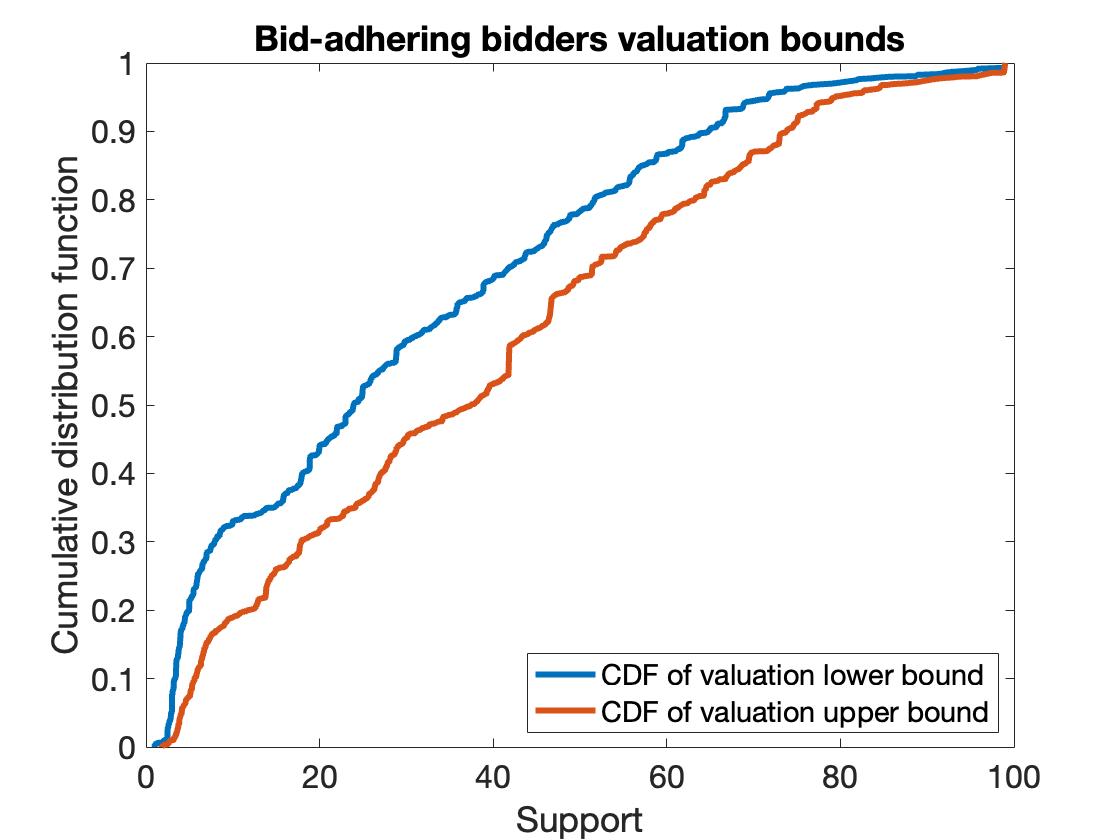}
\caption{\centering CDF of Lower and Upper Bounds for Bid-adhering Bidders Valuation}
	\label{fig:irrational_density}
\end{figure}

Figure \ref{fig:irrational_density} displays the estimated bounds for bid-adhering bidders. The blue line represents the lower bound of the value CDF, and the red line represents the upper bound of the value CDF. 

\newpage 
\subsection{Bid-adhering Bidders Bidding Pattern}\label{app:support_prob}

\begin{figure}[htb!]
\centering
\begin{subfigure}{.5\textwidth}
  \centering
  \includegraphics[width=0.8\linewidth]{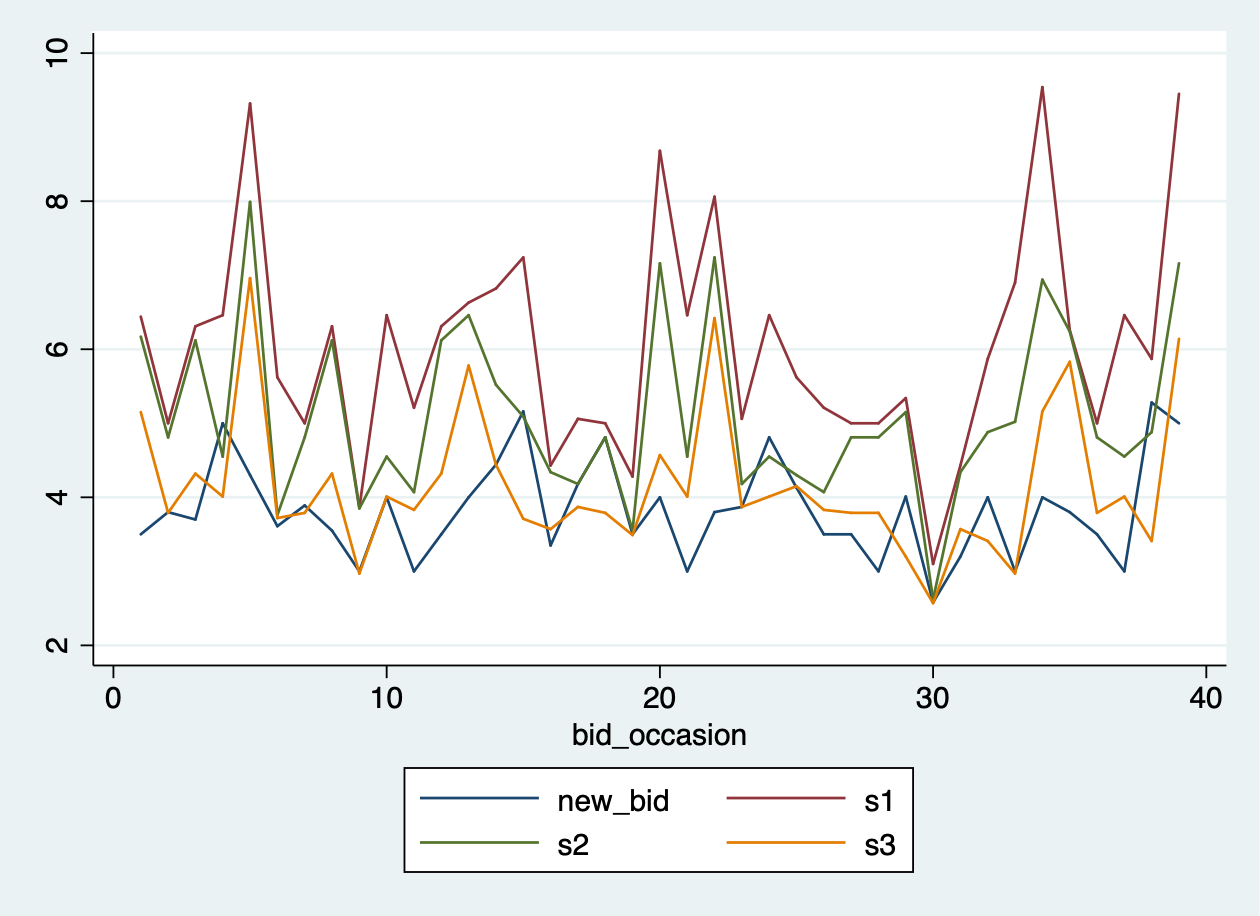}
  \captionsetup{font=footnotesize}
  \caption{\centering Bidding Pattern of Bidder A}
  \footnotesize\textsuperscript{*} Figure \ref{fig:bid_naive_1} displays the bidding pattern of a bid-adhering bidder, referred to as bidder A. Among 37 bid-changing events, bidder A adhered to the bid recommendation 9 times, with a bid-adhering rate of $24.32\%$.
  \label{fig:bid_naive_1}
\end{subfigure}%
\medskip
\begin{subfigure}{.5\textwidth}
  \centering
  \includegraphics[width=0.8\linewidth]{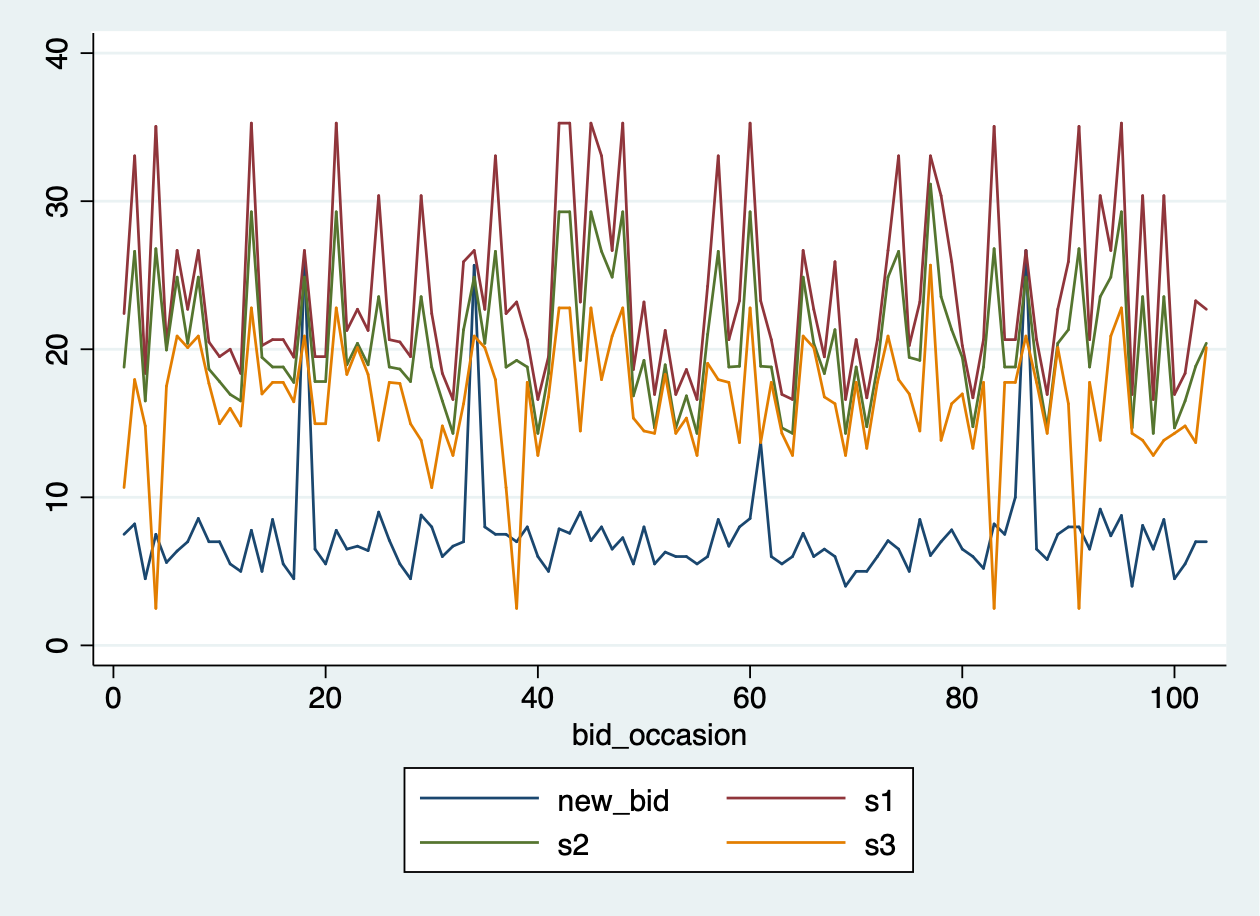}
  \captionsetup{font=footnotesize}
  \caption{Bidding Pattern of Bidder B}
  \footnotesize\textsuperscript{*} Figure \ref{fig:bid_naive_2} displays the bidding pattern of a bid-adhering bidder, referred to as bidder B. Among 103 bid-changing events, bidder B adhered to the bid recommendation 4 times, with a bid-adhering rate of $3.88\%$.
  \label{fig:bid_naive_2}
\end{subfigure}
\caption{\centering Submitted Bids and Observed Bid Recommendations for Bid-adhering Bidders}
\label{fig:bid_naive}
\end{figure}

Figure \ref{fig:bid_naive_1} presents two bidding pattern from bid-adhering bidders, supporting a probabilistic bidding model.

\newpage 
\subsection{Data-driven Polynomial Strategy under Polynomial Bidding Model for Bid-adhering Bidders }
\label{app:bid_adhering_strategy}

\begin{figure}[htb!]
\centering
\begin{subfigure}{.5\textwidth}
  \centering
  \includegraphics[width=.8\linewidth]{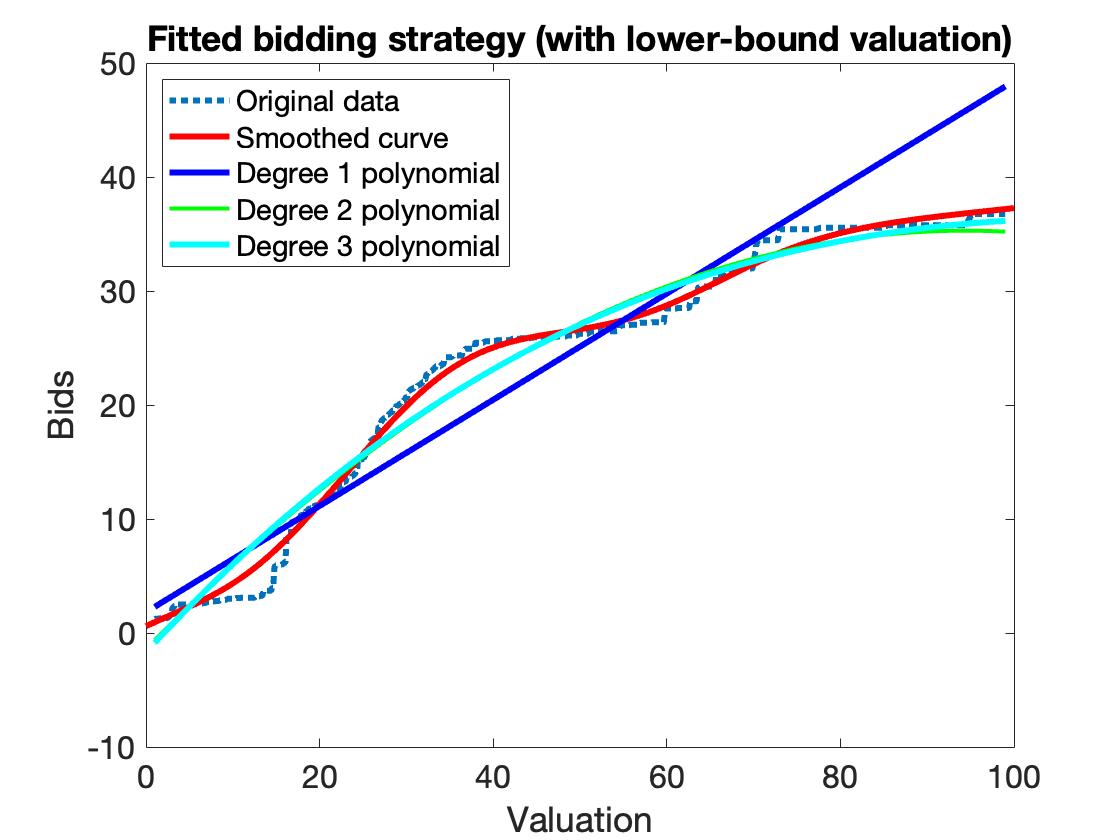}
  \label{fig:bid_lb}
\end{subfigure}%
\begin{subfigure}{.5\textwidth}
  \centering
  \includegraphics[width=.8\linewidth]{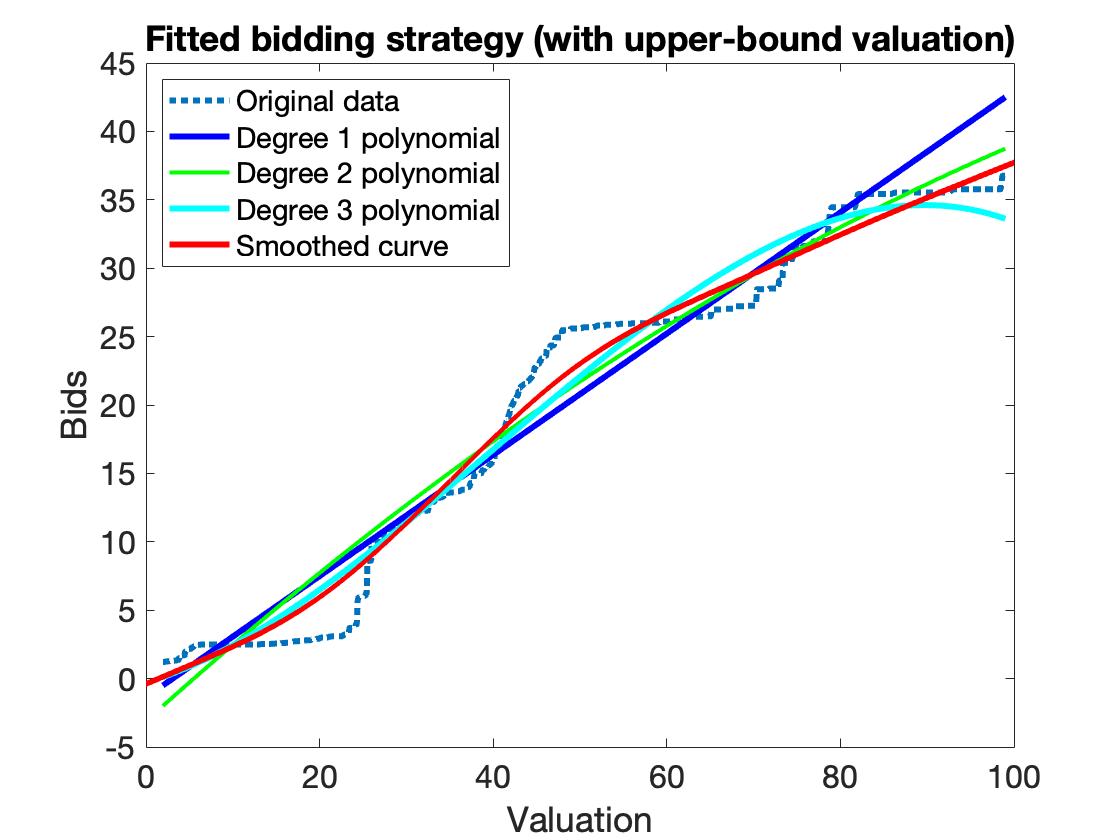}
  \label{fig:bid_ub}
\end{subfigure}
\caption{\centering Fitted polynomial bidding model for bid-adhering bidders}
\label{fig:bid_strategy}
\end{figure}

Figure \ref{fig:bid_strategy} presents the fitted polynomial bidding strategy using both the lower- and upper-bound valuation distributions, along with observed bids from bid-adhering bidders. To estimate this strategy, we generate 10,000 valuation samples from the estimated bounds and match them to bid samples in increasing order, assuming that bidders with higher valuations place higher bids. The polynomial model is fitted separately for both lower- and upper-bound valuations.

\end{document}